\documentclass{llncs}
\pdfoutput=1
\usepackage[utf8]{inputenc}
\usepackage[normalem]{ulem}
\usepackage{url}
\usepackage{xcolor}
\usepackage{amssymb}
\usepackage{amsmath}
\usepackage{paralist}
\usepackage[ruled,vlined,linesnumbered]{algorithm2e}
\usepackage[pdfborder={0 0 0}]{hyperref}
\usepackage{booktabs}
\usepackage{dblfloatfix}
\usepackage{caption}
\usepackage{subcaption}
\usepackage{wrapfig}

\DeclareCaptionLabelFormat{opening}{#2}
\usepackage{tikz}
\usetikzlibrary{calc}
\usetikzlibrary{backgrounds}

\hypersetup{
  colorlinks,
  linkcolor={red!50!black},
  citecolor={green!50!black},
  urlcolor={blue!80!black}
}

\newcommand{\bjparagraph}[1]{\vspace*{4pt}\noindent{\bf #1 }}

\newcommand{\squeeeze}{\vspace*{-6pt}}
\newcommand{\squeeeeze}{\vspace*{-8pt}}

\def\squareforqed{\hbox{\rlap{$\sqcap$}$\sqcup$}}
\def\qed{\ifmmode\squareforqed\else{\unskip\nobreak\hfil
\penalty50\hskip1em\null\nobreak\hfil\squareforqed
\parfillskip=0pt\finalhyphendemerits=0\endgraf}\fi}

\newcommand{\cmplt}[1]{<_{#1}}

\newcommand{\anevent}{e}

\newcommand{\exseq}{\tau}

\newcommand{\mtequiv}{\simeq}

\newcommand{\event}{e}

\newcommand{\happensbefore}[1]{\rightarrow_{#1}}

\newcommand{\uval}{\textrm{$u$}}
\newcommand{\vval}{\textrm{$v$}}
\newcommand{\wval}{\textrm{$w$}}
\newcommand{\rreg}{\textsf{\$}\textrm{$r$}\xspace}
\newcommand{\sreg}{\textsf{\$}\textrm{$s$}\xspace}
\newcommand{\treg}{\textsf{\$}\textrm{$t$}\xspace}
\newcommand{\xvar}{\texttt{\bf{x}}\xspace}
\newcommand{\yvar}{\texttt{\bf{y}}\xspace}
\newcommand{\zvar}{\texttt{\bf{z}}\xspace}
\newcommand{\store}[2]{\textsf{store: }{#1}\texttt{:=}{#2}}
\newcommand{\load}[2]{\textsf{load: }{#1}\texttt{:=}{#2}}
\newcommand{\st}[1]{\textsf{st(}{#1}\textsf{)}}
\newcommand{\ld}[1]{\textsf{ld(}{#1}\textsf{)}}
\newcommand{\upd}[1]{\textsf{u(}{#1}\textsf{)}}
\newcommand{\fence}{\textsf{fence}}
\newcommand{\aprog}{\textrm{$\mathcal{P}$}}
\newcommand{\locstate}{\textrm{$\mathbb{L}$}}
\newcommand{\mem}{\textrm{$\mathbb{M}$}}
\newcommand{\ptid}{\textrm{$p$}\xspace}
\newcommand{\qtid}{\textrm{$q$}\xspace}

\newcommand{\regstate}{\textrm{$\mathbb{R}$}}
\newcommand{\bufstate}{\textrm{$\mathbb{B}$}}
\newcommand{\aconf}{\textrm{$c$}}
\newcommand{\updof}[1]{\textsf{upd}\textrm{$(#1)$}}
\newcommand{\stupdof}[1]{\textsf{upd${}_\textsf{st}$}\textrm{$(#1)$}}
\newcommand{\chronof}[1]{\textrm{$\mathcal{T}_C(#1)$}}
\newcommand{\tsotraceof}[1]{\textrm{$\mathcal{T}(#1)$}}
\newcommand{\roweof}[1]{\textsf{upd${}_\textsf{ld}$(}{#1}\textsf{)}}
\newcommand{\tidof}[1]{\textsf{tid(}{#1}\textsf{)}}

\newcommand{\evtzero}{\textrm{$\anevent^0$}}
\newcommand{\evtinf}{\textrm{$\anevent^\infty$}}
\newcommand{\pc}{\textsf{pc}}
\newcommand{\nat}{\textrm{$\mathbb{N}$}} 
\newcommand{\tids}{\textsf{TID}} 
\newcommand{\auxtids}{\textsf{AuxTID}} 
\newcommand{\memlocs}{\textsf{MemLoc}} 
\newcommand{\events}{\textsf{Event}} 
\newcommand{\vecclocks}{\textsf{VecClocks}} 
\newcommand{\pindent}{\hphantom{MM}}

\newcommand{\porel}{\textrm{$\rightarrow^{\textsf{po}}_{\tau}$}}
\newcommand{\surel}{\textrm{$\rightarrow^{\textsf{su}}_{\tau}$}}
\newcommand{\uurel}{\textrm{$\rightarrow^{\textsf{uu}}_{\tau}$}}
\newcommand{\csrcrel}{\textrm{$\rightarrow^{\textsf{src-ct}}_{\tau}$}}
\newcommand{\ccfrel}{\textrm{$\rightarrow^{\textsf{cf-ct}}_{\tau}$}}
\newcommand{\ufrel}{\textrm{$\rightarrow^{\textsf{uf}}_{\tau}$}}
\newcommand{\strel}{\textrm{$\rightarrow^{\textsf{st}}_{\tau}$}}
\newcommand{\srcrel}{\textrm{$\rightarrow^{\textsf{src-ss}}_{\tau}$}}
\newcommand{\cfrel}{\textrm{$\rightarrow^{\textsf{cf-ss}}_{\tau}$}}
\newcommand{\porelof}[1]{\textrm{$\rightarrow^{\textsf{po}}_{#1}$}}
\newcommand{\surelof}[1]{\textrm{$\rightarrow^{\textsf{su}}_{#1}$}}
\newcommand{\uurelof}[1]{\textrm{$\rightarrow^{\textsf{uu}}_{#1}$}}
\newcommand{\csrcrelof}[1]{\textrm{$\rightarrow^{\textsf{src-ct}}_{#1}$}}
\newcommand{\ccfrelof}[1]{\textrm{$\rightarrow^{\textsf{cf-ct}}_{#1}$}}
\newcommand{\ufrelof}[1]{\textrm{$\rightarrow^{\textsf{uf}}_{#1}$}}
\newcommand{\strelof}[1]{\textrm{$\rightarrow^{\textsf{st}}_{#1}$}}
\newcommand{\srcrelof}[1]{\textrm{$\rightarrow^{\textsf{src-ss}}_{#1}$}}
\newcommand{\cfrelof}[1]{\textrm{$\rightarrow^{\textsf{cf-ss}}_{#1}$}}

\newcommand{\ctrel}{\rightarrow^{\textsf{ct}}_{\tau}}

\renewcommand{\emptyset}{\varnothing}

\newcommand{\etal}{\mbox{\textit{et al.}}\xspace}

\newcommand{\tikzwrapfigbg}{%
  \begin{pgfonlayer}{background}
    \path[fill=gray!20,rounded corners]
    (current bounding box.south west) rectangle
    (current bounding box.north east);
\end{pgfonlayer}}

\begin{document}

\title{Stateless Model Checking for TSO and PSO}

\author{Parosh Abdulla \and Stavros Aronis \and Mohammed Faouzi Atig \and\\ Bengt Jonsson \and Carl Leonardsson \and Konstantinos Sagonas}

\institute{Dept.\ of Information Technology, Uppsala University, Sweden}

\date{\today}

\maketitle

\begin{abstract}
We present a technique for efficient
stateless model checking of programs that execute
under the relaxed memory models TSO and PSO.
The basis for our technique is a novel
representation of executions under TSO and PSO, called
\emph{chronological traces}. Chronological traces induce a
partial order relation on relaxed memory executions, capturing dependencies
that are needed to represent the interaction via shared variables.
They are optimal in the sense that they only distinguish computations that
are inequivalent under the widely-used representation by Shasha and Snir.
This allows an optimal dynamic partial order reduction algorithm to explore
a minimal number of executions while still guaranteeing full coverage.
We apply our techniques to check, under the TSO and PSO memory models,
LLVM assembly produced for C/pthreads programs.
Our experiments show that our technique reduces the verification effort
for relaxed memory models to be almost that for the standard model of
sequential consistency.
In many cases, our implementation significantly outperforms other
comparable tools.
\end{abstract}

\section{Introduction}

Verification and testing of concurrent programs is difficult, since one must
consider all the different ways in which instructions of different
threads can be interleaved.
To make matters worse, most architectures implement
\emph{relaxed memory models},
such as TSO and PSO~\cite{sparc-v9-manual,adve1996shared},
which make threads interact in even more and subtler ways than by
standard interleaving.
For example, a processor may reorder loads and stores by the same thread
if they target different addresses, or it may buffer stores in a local queue.

A successful technique for finding concurrency bugs (i.e., defects that
arise only under some thread schedulings), and for
verifying their absence, is
\emph{stateless model checking} (SMC)~\cite{Godefroid:popl97}, also known as
\emph{systematic concurrency testing}~\cite{LC:reachabilitytesting,WSG:icse11}.
Starting from a test, i.e., a way to run a program and obtain some
expected result, which is terminating and threadwisely deterministic
(e.g. no data-nondeterminism), SMC systematically explores the set of
all thread schedulings that are possible during runs of this test.
A
special runtime scheduler drives the SMC exploration by making decisions
on scheduling whenever such decisions may affect the interaction
between threads, so that the exploration covers all possible
executions and detects any unexpected test results, program crashes,
or assertion violations. The technique is completely automatic, has no
false positives, does not suffer from memory explosion, and can easily
reproduce the concurrency bugs it detects. SMC has been successfully
implemented in tools such as VeriSoft~\cite{Godefroid:verisoft-journal},
\textsc{Chess}~\cite{MQBBNN:chess}, and Concuerror~\cite{Concuerror:ICST13}.

There are two main problems for using SMC in programs that run under
relaxed memory models (RMM).
The first problem is that
already under the standard model of {\em sequential consistency}
(SC)
the number of possible thread schedulings grows exponentially with
the length of program execution.
This problem has been addressed by \emph{partial order reduction} (POR)
techniques that
achieve coverage of {\em all} thread schedulings,
by exploring only a representative
subset~\cite{Valmari:reduced:state-space,Peled:representatives,Godefroid:thesis,CGMP:partialorder}.
POR has been adapted to SMC in the form of
{\em Dynamic Partial Order Reduction} (DPOR)~\cite{FG:dpor}, which has been
further developed in recent years~\cite{SeAg:haifa06,LC:reachabilitytesting,LKMA:fase10,SKH:acsd12,TKLLMA:forte12,abdulla2014optimal}.
DPOR is based on augmenting each execution by
a {\em happens-before relation}, which is a partial order that captures
dependencies between operations of the threads.
Two executions can be regarded as equivalent if they induce
the same happens-before relation, and it is therefore sufficient to
explore one execution in each equivalence class
(called a \emph{Mazurkiewicz~trace}~\cite{Mazurkiewicz:traces}).
DPOR algorithms guarantee to explore at least one execution in each
equivalence class, thus attaining full coverage with reduced cost.
A recent optimal algorithm~\cite{abdulla2014optimal} guarantees to
explore \emph{exactly} one execution per equivalence class.

The second problem is that
in order to extend SMC to handle relaxed memory models, the operational semantics of programs must be extended to represent the effects of RMM.
The natural approach is to augment the program state with additional structures,
e.g., store buffers in the case of TSO, that model the effects of
RMM~\cite{abdulla2012counter,AlKNT13,park1995executable}.
This causes
blow-ups in the number of possible executions, in addition to those possible
under SC. However, most of these additional executions are equivalent to some
SC execution. To efficiently apply SMC to handle RMM, we must therefore 
extend DPOR to avoid redundant exploration of equivalent executions.
The natural definition of ``equivalent'' under RMM
can be derived from the abstract representation of executions due to
Shasha and Snir~\cite{shasha1988efficient},
here called {\em Shasha-Snir traces},
which is often used
in model checking and runtime verification~\cite{DBLP:journals/jpdc/KrishnamurthyY96,DBLP:journals/tc/LeeP01,BM08,BurnimSS11,AlglaveM11,BouajjaniDM13}.
Shasha-Snir traces consist of an ordering relation between dependent operations,
which generalizes the standard happens-before relation on SC executions;
indeed, under SC, the equivalence relation induced by Shasha-Snir traces
coincides with Mazurkiewicz traces.
It would thus be natural to base DPOR for RMM
on the happens-before relation induced by Shasha-Snir traces.
However, this relation is in general cyclic
(due to reorderings possible under RMM) and can therefore
not be used as a basis for DPOR (since it is not a partial order).
To develop an efficient technique for SMC under RMM we therefore need
to find a different representation of executions under RMM. The
representation should define an acyclic happens-before relation. Also,
the induced trace equivalence should coincide with the equivalence
induced by Shasha-Snir traces.

\bjparagraph{Contribution}
In this paper, we show how to apply SMC to TSO and PSO in a way
that achieves maximal possible reduction using DPOR, in the sense that
redundant exploration of equivalent executions is avoided.
A cornerstone in our contribution is a novel
representation of executions under RMM, called
{\em chronological traces}, which define a
happens-before relation on the events in a carefully designed
representation of program executions.
Chronological traces are a succinct
canonical representation of executions, in the sense that
there is a one-to-one correspondence between chronological traces
and Shasha-Snir traces.
Furthermore, the happens-before relation induced by chronological traces is
a partial order, and can therefore be used as a basis for DPOR.
In particular, the Optimal-DPOR algorithm of~\cite{abdulla2014optimal} will
explore exactly one execution per Shasha-Snir trace.
In particular, for so-called {\em robust} programs that
are not affected by RMM (these include data-race-free programs), 
Optimal-DPOR will explore as many executions under RMM as under SC:
this follows from the one-to-one correspondence between chronological traces and
Mazurkiewicz traces under SC.
Furthermore, robustness can itself be considered
a correctness criterion, which can also be automatically checked with our
method (by checking whether the number of equivalence classes is increased
when going from SC to RMM).

We show the power of our technique by using it to implement an
efficient stateless model checker, which for C programs with
pthreads explores all executions of
a test-case or a program, up to some bounded length.
During exploration of an execution, our implementation generates
the corresponding chronological trace.
Our implementation employs
the source-DPOR algorithm~\cite{abdulla2014optimal}, which
is simpler than Optimal-DPOR, but about equally effective.
Our experimental results for analyses under SC, TSO and PSO of
number of intensely racy benchmarks and programs written in C/pthreads, shows
that
\begin{inparaenum}[(i)]
\item
the effort for verification under TSO and PSO is not much larger than the
effort for verification under SC, and
\item
our implementation compares favourably against CBMC,
a state-of-the-art bounded model checking tool,
showing the potential of our approach.
\end{inparaenum}

\newcommand{\wrapfigtopfix}{\vspace{-2.2em}}
\newcommand{\wrapfigbotfix}{\vspace{-0.8em}}

\section{Overview of Main Concepts}
\label{sec:intuition}

This section informally motivates and explains the main concepts of the paper.
To focus the presentation, we consider mainly the TSO model.
TSO is relevant because it is implemented in the widely used
x86 as well as SPARC architectures.
We first introduce TSO and its semantics. Thereafter we introduce Shasha-Snir
traces, which abstractly represent the orderings between dependent events
in an execution. Since Shasha-Snir traces are cyclic, we introduce an
extended representation of executions, for which a natural happens-before
relation is acyclic. We then describe how this happens-before relation
introduces undesirable distinctions between executions, and how our
new representation of chronological traces remove these distinctions. Finally,
we illustrate how a DPOR algorithm exploits the happens-before relation
induced by chronological traces to explore only a minimal number of
executions, while still guaranteeing full coverage.

\begin{wrapfigure}[7]{r}{.395\textwidth}
  \wrapfigtopfix
  \centering
  \begin{tikzpicture}
    \node (wx1) at (0,0) [] {
      \small{\store{\xvar\vphantom{\yvar}}{1}}
    };
    \node (ry) at ($(wx1.west)+(0,-0.5)$) [anchor=west] {
      \small{\load{\rreg}{\yvar}}
    };

    \node (wy1) at (1.85,0) [] {
      \small{\store{\yvar}{1}}
    };
    \node (rx) at ($(wy1.west)+(0,-0.5)$) [anchor=west] {
      \small{\load{\sreg}{\xvar\vphantom{\yvar}}}
    };

    \node (p) at ($(wx1)+(0,.5)$) [] {\small{\ptid}};
    \node (p) at ($(wy1)+(0,.5)$) [] {\small{\qtid}};
    \tikzwrapfigbg
  \end{tikzpicture}
  \wrapfigbotfix
  \caption{A program implementing the classic idiom of Dekker's mutual
    exclusion algorithm.}\label{fig:dekker}
\label{fig:set:creation}
\end{wrapfigure}

\bjparagraph{TSO --- an Introduction}
TSO relaxes the ordering between stores and subsequent loads to
different memory locations.
This can be explained  operationally by equipping each
thread with a {\em store buffer}~\cite{sewell2010x86},
which is a FIFO queue that contains pending store
operations. When a thread executes a store instruction, the store does
not immediately affect memory. Instead it is delayed and enqueued in
the store buffer. Nondeterministically, at some later point an {\em
  update} event occurs, dequeueing the oldest store from the store
buffer and updating the memory correspondingly.
Load instructions take effect immediately, without being delayed.
Usually a load reads a value from memory. However, if the store buffer of
the same thread contains a store to the same memory location, the
value is instead taken from the store in the store buffer.

\begin{wrapfigure}[9]{r}{.49\textwidth}
  \centering
  \begin{tikzpicture}
    \node (wx1) at (0,0) [] {
      \small{\ptid: \store{\xvar\vphantom{\yvar}}{1} \hspace{5pt}\texttt{// Enqueue store}}
    };
    \node (ry) at ($(wx1.west)+(0,-0.4)$) [anchor=west] {
      \small{\ptid: \load{\rreg}{\yvar} \hspace{5pt}\texttt{// Load value 0}}
    };
    \node (wy1) at ($(ry.west)+(0,-0.4)$) [anchor=west] {
      \small{\pindent\qtid: \store{\yvar}{1} \hspace{1em}\texttt{// Enqueue store}}
    };
    \node (uy1) at ($(wy1.west)+(0,-0.4)$) [anchor=west] {
      \small{\pindent\qtid: \textsf{update} \hspace{2.7em}\texttt{// $\yvar = 1$ in memory}}
    };
    \node (rx) at ($(uy1.west)+(0,-0.4)$) [anchor=west] {
      \small{\pindent\qtid: \load{\sreg}{\xvar\vphantom{\yvar}} \hspace{5pt}\texttt{// Load value 0}}
    };
    \node (uy1) at ($(rx.west)+(0,-0.4)$) [anchor=west] {
      \small{\ptid: \textsf{update} \hspace{2.5em}\texttt{// $\xvar = 1$ in memory}}
        };
    \tikzwrapfigbg
  \end{tikzpicture}
  \wrapfigbotfix
  \caption{An execution of the program in \figurename~\ref{fig:dekker}.
    Notice that $\rreg = \sreg = 0$ at the end.}\label{fig:dekker:cmp}
\end{wrapfigure}

To see why this buffering semantics may cause unexpected program
behaviors, consider the small program in \figurename~\ref{fig:dekker}.
It consists of two threads \ptid and \qtid. The thread \ptid first
stores 1 to the memory location \xvar, and then loads the value at
memory location \yvar into its register \rreg. The thread \qtid is
similar. All memory locations and registers are assumed to have
initial values~0.
It is easy to see that  under the SC semantics, it is
impossible for the program to terminate in a state where both registers
\rreg and \sreg hold the value~0.
However, under the buffering semantics of TSO, such a final state is
possible.
\figurename~\ref{fig:dekker:cmp} shows one such program execution.
We see that the store to \xvar happens at the beginning of the execution,
but does not take effect with respect to memory until the very end
of the execution. Thus the store to \xvar and the load to \yvar appear
to take effect in an order opposite to how they occur in the program
code. This allows the execution to terminate with
$\rreg = \sreg = 0$.

\bjparagraph{Shasha-Snir Traces for TSO}
\label{sect:tso-hb}
Partial order reduction is based on the idea of capturing the possible
orderings between dependent operations of different threads by means of a
happens-before relation.
When threads interact via shared variables,
two instructions
are considered dependent if they access the same global variable,
and at least one is a write.
For relaxed memory models, Shasha and Snir~\cite{shasha1988efficient}
introduced an abstract representation of executions, here
referred to as {\em Shasha-Snir traces}, which
captures such dependencies in a natural way.
Shasha-Snir traces induce equivalence classes of executions. Under
sequential consistency, those classes coincide with the Mazurkiewicz
traces. Under a relaxed memory model, there are also additional
Shasha-Snir traces corresponding to the non-sequentially consistent
executions.

\begin{wrapfigure}[8]{r}{.41\textwidth}
  \wrapfigtopfix
  \centering
  \begin{tikzpicture}
    \node (wx1) at (0,0) [] {
      \small{\store{\xvar\vphantom{\yvar}}{1}}
    };
    \node (ry) at ($(wx1.west)+(0,-1)$) [anchor=west] {
      \small{\load{\rreg}{\yvar}}
    };

    \node (wy1) at (2.5,0) [] {
      \small{\store{\yvar}{1}}
    };
    \node (rx) at ($(wy1.west)+(0,-1)$) [anchor=west] {
      \small{\load{\sreg}{\xvar\vphantom{\yvar}}}
    };

    \node (p) at ($(wx1)+(0,.5)$) [] {\small{\ptid}};
    \node (p) at ($(wy1)+(0,.5)$) [] {\small{\qtid}};

    \draw[->,line width=1pt] (wx1) -- (wx1 |- ry.north);
    \draw[->,line width=1pt] (wy1) -- (wy1 |- rx.north);
    \draw[->,line width=1pt,out=0,in=180] (ry) to (wy1);
    \draw[->,line width=1pt,out=180,in=0] (rx) to (wx1);
    \tikzwrapfigbg
  \end{tikzpicture}
  \wrapfigbotfix
  \caption{The Shasha-Snir trace corresponding to the execution in
    \figurename~\ref{fig:dekker:cmp}.}\label{fig:dekker:cmp:trace}
\end{wrapfigure}
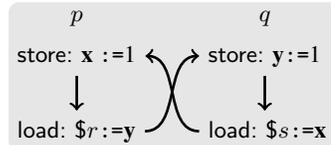

A Shasha-Snir trace is a directed graph, where edges capture
observed event orderings.
The nodes in a Shasha-Snir trace are the executed instructions.
For each thread, there are edges between each pair of subsequent instructions,
creating a total order for each thread.
For
two instructions $i$ and $j$ in different
threads, there is an edge $i\rightarrow j$ in a trace when $i$ causally
precedes $j$.
This happens when $j$ reads a value that was written by $i$, when
$i$ reads a memory location
that is subsequently updated by $j$, or when $i$ and $j$
are subsequent writes to the same memory location.
In \figurename~\ref{fig:dekker:cmp:trace} we show the Shasha-Snir trace
for the execution in \figurename~\ref{fig:dekker:cmp}.

\bjparagraph{Making the Happens-Before Relation Acyclic}
Shasha-Snir traces naturally represent the
dependencies between operations in an execution, and are therefore
a natural basis for applying DPOR.
However, a major problem is that the happens-before relation induced by
the edges is in general cyclic, and thus not a partial order. This
can be seen already in the graph in \figurename~\ref{fig:dekker:cmp:trace}.
This problem can be addressed by adding nodes that represent explicit update
events. That would be natural since such events occur in the representation
of  the execution in \figurename~\ref{fig:dekker:cmp}.
When we consider the edges of the Shasha-Snir trace, we observe that
although there is a conflict between $\ptid:
\load{\rreg}{\yvar}$ and $\qtid: \store{\yvar}{1}$, swapping
their order in the execution in \figurename~\ref{fig:dekker:cmp} has
no observable effect; the load still gets the same value from memory.
Therefore, we should only be concerned with the order of the load relative
to the update event $\qtid: \textsf{update}$.

\begin{wrapfigure}[10]{r}{.5\textwidth}
  \wrapfigtopfix
  \centering
  \begin{tikzpicture}[scale=0.82]
    \node (wx1) at (0,0) [] {
      \small{\store{\xvar\vphantom{\yvar}}{1}}
    };
    \node (ry) at ($(wx1.west)+(0,-2)$) [anchor=west] {
      \small{\load{\rreg}{\yvar}}
    };
    \node (ux1) at ($(wx1)+(.5,-1)$) [anchor=west] {
      \small{\textsf{update}}
    };

    \node (wy1) at (5,0) [] {
      \small{\store{\yvar}{1}}
    };
    \node (rx) at ($(wy1.west)+(0,-2)$) [anchor=west] {
      \small{\load{\sreg}{\xvar\vphantom{\yvar}}}
    };
    \node (uy1) at ($(wy1)+(-.5,-1)$) [anchor=east] {
      \small{\textsf{update}}
    };

    \node (p) at ($(wx1)+(0,.5)$) [] {\ptid};
    \node (p) at ($(wy1)+(0,.5)$) [] {\qtid};

    \draw[->,line width=1pt] (wx1) -- (wx1 |- ry.north);
    \draw[->,line width=1pt] (wy1) -- (wy1 |- rx.north);
    \draw[->,line width=1pt] (wx1) -- (ux1);
    \draw[->,line width=1pt] (wy1) -- (uy1);
    \draw[->,line width=1pt,out=0,in=180] (ry) to (uy1);
    \draw[->,line width=1pt,out=180,in=0] (rx) to (ux1);
    \tikzwrapfigbg
  \end{tikzpicture}
  \wrapfigbotfix
  \caption{A trace for the execution in
    \figurename~\ref{fig:dekker:cmp} where updates are separated
    from stores.}\label{fig:dekker:ctrace}
\end{wrapfigure}
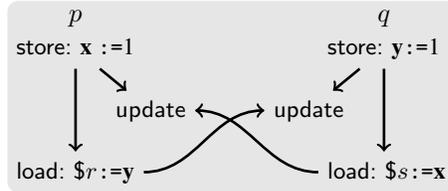

These observations suggest to define a representation of traces
that separates stores from updates.
In \figurename~\ref{fig:dekker:ctrace} we have redrawn the trace from
\figurename~\ref{fig:dekker:cmp:trace}.
Updates are separated from stores, and we order updates, rather than
stores, with operations of other threads. Thus, there are edges between
updates to and loads from the same memory location,
and between two updates to the same memory location.
In \figurename~\ref{fig:dekker:ctrace}, there is an
edge from each store to the corresponding update, reflecting the principle that
the update cannot occur before the store. There are
edges between loads and updates of the same memory location,
reflecting that swapping their order will affect the observed
values. However, notice that for this program
there are no edges between the updates and
loads of the same thread, since they access different
memory locations.

\bjparagraph{Chronological Traces for TSO}
\label{sect:tso-ct}
Although the new representation is a valid partial order, it will in
many cases distinguish executions that are semantically equivalent according
to the Shasha-Snir traces.
The reason for this is the mechanism of TSO buffer
forwarding: When a thread executes a load to a memory location
\xvar, it will first check its store buffer. If the buffer contains
a store to \xvar, then the load returns the value of the newest such
store buffer entry instead of loading the value from memory.
This causes difficulties for a happens-before relation that
orders any update with any load of the same memory location.

\begin{wrapfigure}[6]{r}{.33\textwidth}
  \wrapfigtopfix
  \centering
  \begin{tikzpicture}
    \node (wx1) at (0,0) [] {
      \small{\store{\xvar}{1}}
    };
    \node (rx) at ($(wx1.west)+(0,-0.5)$) [anchor=west] {
      \small{\load{\rreg}{\xvar}}
    };

    \node (wx2) at (1.75,0) [] {
      \small{\store{\xvar}{2}}
    };

    \node (p) at ($(wx1)+(0,.5)$) [] {\small{\ptid}};
    \node (q) at ($(wx2)+(0,.5)$) [] {\small{\qtid}};
    \tikzwrapfigbg
  \end{tikzpicture}
  \wrapfigbotfix
  \caption{A program illustrating buffer forwarding.}\label{fig:rowe:prog}
\end{wrapfigure}
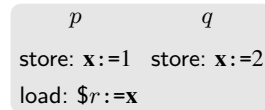
\begin{figure}[h]
  \centering
    \begin{subfigure}[b]{.32\linewidth}
      \begin{tikzpicture}
        \node (wx1) at (0,0) [] {
          \small{\store{\xvar}{1}}
        };
        \node (rx) at ($(wx1.west)+(1,-.7)$) [anchor=west] {
          \small{\load{\rreg}{\xvar}}
        };
        \node (ux1) at ($(wx1) + (0,-2)$) [] {
          \small{\textsf{update}}
        };

        \node (wx2) at (2.1,0) [] {
          \small{\store{\xvar}{2}}
        };
        \node (ux2) at ($(wx2) + (0,-1.5)$) [] {
          \small{\textsf{update}}
        };

        \draw[->,line width=1pt] (wx1) -- (rx);
        \draw[->,line width=1pt] (wx1) -- (ux1);
        \draw[->,line width=1pt] (wx2) -- (ux2);

        \draw[->,line width=1pt] (ux2) -- (ux1);
        \draw[->,line width=1pt] (rx) -- (ux2);

        \node (p) at ($(wx1)+(0,.5)$) [] {\small{\ptid}};
        \node (q) at ($(wx2)+(0,.5)$) [] {\small{\qtid}};
      \tikzwrapfigbg
      \end{tikzpicture}
      \caption{}\label{fig:rowe:redundant:a}
    \end{subfigure}
    \begin{subfigure}[b]{.32\linewidth}
      \begin{tikzpicture}
        \node (wx1) at (0,0) [] {
          \small{\store{\xvar}{1}}
        };
        \node (rx) at ($(wx1.west)+(1,-1.3)$) [anchor=west] {
          \small{\load{\rreg}{\xvar}}
        };
        \node (ux1) at ($(wx1) + (0,-2)$) [] {
          \small{\textsf{update}}
        };

        \node (wx2) at (2.1,0) [] {
          \small{\store{\xvar}{2}}
        };
        \node (ux2) at ($(wx2) + (0,-.7)$) [] {
          \small{\textsf{update}}
        };

        \draw[->,line width=1pt] (wx1) -- (rx);
        \draw[->,line width=1pt] (wx1) -- (ux1);
        \draw[->,line width=1pt] (wx2) -- (ux2);

        \draw[->,line width=1pt] (ux2) -- (rx);
        \draw[->,line width=1pt] (rx) -- (ux1);

        \node (p) at ($(wx1)+(0,.5)$) [] {\small{\ptid}};
        \node (q) at ($(wx2)+(0,.5)$) [] {\small{\qtid}};
      \tikzwrapfigbg
      \end{tikzpicture}
      \caption{}\label{fig:rowe:redundant:b}
    \end{subfigure}
    \begin{subfigure}[b]{.32\linewidth}
      \begin{tikzpicture}
        \node (wx1) at (0,0) [] {
          \small{\store{\xvar}{1}}
        };
        \node (rx) at ($(wx1.west)+(0,-2)$) [anchor=west] {
          \small{\load{\rreg}{\xvar}}
        };
        \node (ux1) at ($(wx1) + (1,-1.35)$) [] {
          \small{\textsf{update}}
        };

        \node (wx2) at (2.1,0) [] {
          \small{\store{\xvar}{2}}
        };
        \node (ux2) at ($(wx2) + (0,-.7)$) [] {
          \small{\textsf{update}}
        };

        \draw[->,line width=1pt] (wx1) -- (rx);
        \draw[->,line width=1pt] (wx1) -- (ux1);
        \draw[->,line width=1pt] (wx2) -- (ux2);

        \draw[->,line width=1pt] (ux2) -- (ux1);
        \draw[->,line width=1pt] (ux1) -- (rx);

        \node (p) at ($(wx1)+(0,.5)$) [] {\small{\ptid}};
        \node (q) at ($(wx2)+(0,.5)$) [] {\small{\qtid}};
      \tikzwrapfigbg
      \end{tikzpicture}
      \caption{}\label{fig:rowe:redundant:c}
    \end{subfigure}
  \vspace*{-0.5em}
  \caption{
    Three redundant happens-before relations for
    \figurename~\ref{fig:rowe:prog}.}
  \label{fig:rowe:redundant}
\end{figure}
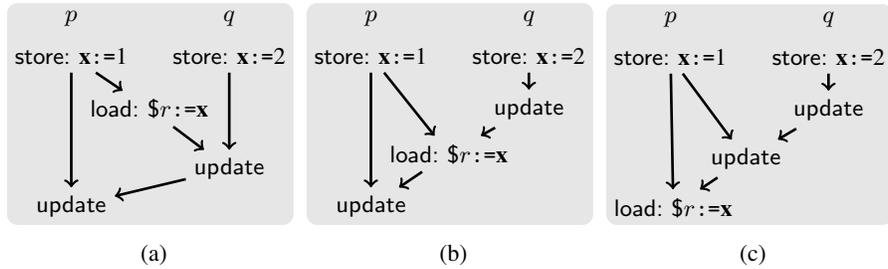
\begin{figure}
  \centering
  \begin{subfigure}[t]{.45\linewidth}
  \centering
  \begin{tikzpicture}
    \node (w0) at (0,0) [] {
      \small{\store{\xvar}{1}}
    };
    \node (r0) at ($(w0.west)+(0,-1.5)$) [anchor=west] {
      \small{\load{\rreg}{\xvar}}
    };
    \node (w1) at (2,0) [] {
      \small{\store{\xvar}{2}}
    };

    \node (p) at ($(w0)+(0,.5)$) [] {\small{\ptid}};
    \node (q) at ($(w1)+(0,.5)$) [] {\small{\qtid}};

    \draw[->,line width=1pt] (w0) -- (r0);
    \draw[->,line width=1pt] (w1) -- (w0);
    \tikzwrapfigbg
  \end{tikzpicture}
  \caption{A Shasha-Snir trace corresponding\\ to all three traces of
    \figurename~\ref{fig:rowe:redundant}.}\label{fig:rowe:tsotrace}
  \end{subfigure}
  \begin{subfigure}[t]{.45\linewidth}
    \centering
    \begin{tikzpicture}
      \node (wx1) at (0,0) [] {
        \small{\store{\xvar}{1}}
      };
      \node (rx) at ($(wx1.west)+(1,-.7)$) [anchor=west] {
        \small{\load{\rreg}{\xvar}}
      };
      \node (ux1) at ($(wx1) + (0,-1.5)$) [] {
        \small{\textsf{update}}
      };

      \node (wx2) at (2.25,0) [] {
        \small{\store{\xvar}{2}}
      };
      \node (ux2) at ($(wx2) + (0,-1.1)$) [] {
        \small{\textsf{update}}
      };

      \draw[->,line width=1pt] (wx1) -- (rx);
      \draw[->,line width=1pt] (wx1) -- (ux1);
      \draw[->,line width=1pt] (wx2) -- (ux2);

      \draw[->,line width=1pt] (ux2) -- (ux1);

      \node (p) at ($(wx1)+(0,.5)$) [] {\small{\ptid}};
      \node (q) at ($(wx2)+(0,.5)$) [] {\small{\qtid}};
    \tikzwrapfigbg
    \end{tikzpicture}
  \caption{The three traces can be merged into this single
    trace.}\label{fig:rowe:merged}
  \end{subfigure}
  \caption{Traces that capture all three
    \figurename~\ref{fig:rowe:redundant:a}, \ref{fig:rowe:redundant:b} and
    \ref{fig:rowe:redundant:c}.}
\end{figure}
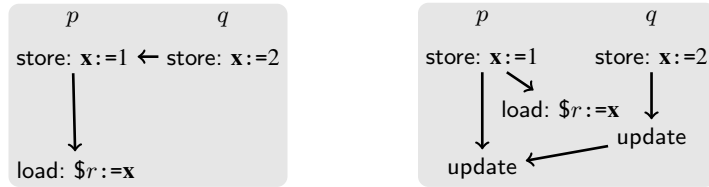

For example, consider the program shown in
\figurename~\ref{fig:rowe:prog}. Any execution of this program will have
two updates and one load to \xvar. Those accesses can be permuted in
six different ways. \figurename~\ref{fig:rowe:redundant:a},
\ref{fig:rowe:redundant:b} and~\ref{fig:rowe:redundant:c} show three
of the corresponding happens-before relations.
In \figurename~\ref{fig:rowe:redundant:a} and \ref{fig:rowe:redundant:b}
the load is satisfied by buffer forwarding, and in
\ref{fig:rowe:redundant:c} by a read from memory.
These three relations all correspond to the same Shasha-Snir trace,
shown in \figurename~\ref{fig:rowe:tsotrace},
and they all have the same observable behavior,
since the value of the load is obtained from the same store.
Hence, we should find a representation of executions that does not distinguish
between these three cases.

We can now describe \emph{chronological traces},
our representation which solves the above problems,
by omitting some of the edges, leaving some nodes unrelated.
More precisely, edges between loads and updates should be
omitted in the following  cases.
\begin{enumerate}
\item
A load is never related to an update originating in
the same thread. This captures the intuition that swapping the order
of such a load and update has no effect other than changing a load
from memory into a load of the same value from buffer, as seen when
comparing \figurename~\ref{fig:rowe:redundant:b} and~\ref{fig:rowe:redundant:c}.
\item
A load \textsf{ld} from a memory location \xvar by a thread \ptid
is never related
to an update by an another thread \qtid, if the update by \qtid
precedes some update to \xvar originating in a store
by \ptid that precedes \textsf{ld}.
This is because the value
written by the update of \qtid is effectively hidden to the load \textsf{ld}
by the update to \xvar by \ptid.
Thus, when we compare \figurename~\ref{fig:rowe:redundant:a} and
\ref{fig:rowe:redundant:b}, we see that the order between the update
by \qtid and the load is irrelevant, since the update by \qtid is
hidden by the update by \ptid (note that the update by \ptid originates in
a store that precedes the load).
\end{enumerate}

When we apply these rules to the example of
\figurename~\ref{fig:rowe:prog}, all of the three representations in
\figurename~\ref{fig:rowe:redundant:a},~\ref{fig:rowe:redundant:b}, and~\ref{fig:rowe:redundant:c}
merge into a single representation shown in \figurename~\ref{fig:rowe:merged}.
In total, we reduce the number of distinguished cases for the program
from six to three.
This is indeed the minimal number of cases that must be distinguished by any
representation, since the different cases result in different
values being loaded by the load instruction or different values in memory at
the end of the execution. Our proposed representation
is optimal for the programs in \figurename~\ref{fig:dekker} and~\ref{fig:rowe:prog}.
In Theorem~\ref{thm:equivalence} of Section~\ref{sec:formalization}
we will show that such an optimality result holds in general.

\bjparagraph{Chronological Traces for PSO}
The TSO and PSO memory models are very similar. The difference is that
PSO does not enforce program order between stores by the same thread
to different memory locations.
To capture this, chronological traces are constructed differently
under TSO and PSO. In particular, under TSO there will always be edges
between all updates of the same thread, but under PSO we omit those
edges when the updates access different memory locations.
In Appendix~\ref{sec:pso} we describe in detail how to adapt the chronological
traces described above to the PSO memory model.

\bjparagraph{DPOR Based on Chronological Traces}
\label{sect:dpor-ct}
Here, we illustrate how
stateless model checking performs DPOR based on chronological
traces, in order to explore one execution per chronological trace.
As example, we use the small program of \figurename~\ref{fig:rowe:prog}.

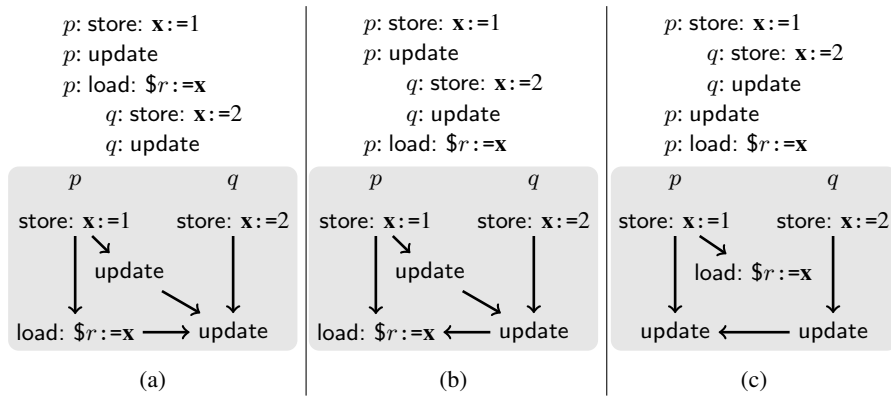
\begin{figure}
  \centering
  \begin{subfigure}[t]{.32\linewidth}
    \centering
    \begin{tikzpicture}
      \node (pw) at (0,0) [] {\small{\ptid: \store{\xvar}{1}}};
      \node (pu) at ($(pw.west)+(0,-.4)$) [anchor=west] {\small{\ptid: \textsf{update}}};
      \node (pr) at ($(pu.west)+(0,-.4)$) [anchor=west] {\small{\ptid: \load{\rreg}{\xvar}}};
      \node (qw) at ($(pr.west)+(0,-.4)$) [anchor=west] {\small{\pindent\qtid: \store{\xvar}{2}}};
      \node (qu) at ($(qw.west)+(0,-.4)$) [anchor=west] {\small{\pindent\qtid: \textsf{update}}};
    \end{tikzpicture}\\
    \begin{tikzpicture}
      \node (wx1) at (0,0) [] {
        \small{\store{\xvar}{1}}
      };
      \node (rx) at ($(wx1) + (0,-1.5)$) {
        \small{\load{\rreg}{\xvar}}
      };
      \node (ux1) at ($(wx1.west)+(1,-.7)$) [] [anchor=west] {
        \small{\textsf{update}}
      };

      \node (wx2) at (2.1,0) [] {
        \small{\store{\xvar}{2}}
      };
      \node (ux2) at ($(wx2) + (0,-1.5)$) [] {
        \small{\textsf{update}}
      };

      \draw[->,line width=1pt] (wx1) -- (rx);
      \draw[->,line width=1pt] (wx1) -- (ux1);
      \draw[->,line width=1pt] (wx2) -- (ux2);

      \draw[->,line width=1pt] (rx) -- (ux2);
      \draw[->,line width=1pt] (ux1) -- (ux2);

      \node (p) at ($(wx1)+(0,.5)$) [] {\small{\ptid}};
      \node (q) at ($(wx2)+(0,.5)$) [] {\small{\qtid}};
      \tikzwrapfigbg
    \end{tikzpicture}
    \caption{}\label{fig:smcrun:a}
  \end{subfigure}
  \vrule
  \begin{subfigure}[t]{.32\linewidth}
    \centering
    \begin{tikzpicture}
      \node (pw) at (0,0) [] {\small{\ptid: \store{\xvar}{1}}};
      \node (pu) at ($(pw.west)+(0,-.4)$) [anchor=west] {\small{\ptid: \textsf{update}}};
      \node (qw) at ($(pu.west)+(0,-.4)$) [anchor=west] {\small{\pindent\qtid: \store{\xvar}{2}}};
      \node (qu) at ($(qw.west)+(0,-.4)$) [anchor=west] {\small{\pindent\qtid: \textsf{update}}};
      \node (pr) at ($(qu.west)+(0,-.4)$) [anchor=west] {\small{\ptid: \load{\rreg}{\xvar}}};
    \end{tikzpicture}\\
    \begin{tikzpicture}
      \node (wx1) at (0,0) [] {
        \small{\store{\xvar}{1}}
      };
      \node (rx) at ($(wx1) + (0,-1.5)$) [] {
        \small{\load{\rreg}{\xvar}}
      };
      \node (ux1) at ($(wx1.west)+(1,-.7)$) [anchor=west] {
        \small{\textsf{update}}
      };

      \node (wx2) at (2.1,0) [] {
        \small{\store{\xvar}{2}}
      };
      \node (ux2) at ($(wx2) + (0,-1.5)$) [] {
        \small{\textsf{update}}
      };

      \draw[->,line width=1pt] (wx1) -- (rx);
      \draw[->,line width=1pt] (wx1) -- (ux1);
      \draw[->,line width=1pt] (wx2) -- (ux2);

      \draw[->,line width=1pt] (ux1) -- (ux2);
      \draw[->,line width=1pt] (ux2) -- (rx);

      \node (p) at ($(wx1)+(0,.5)$) [] {\small{\ptid}};
      \node (q) at ($(wx2)+(0,.5)$) [] {\small{\qtid}};
      \tikzwrapfigbg
    \end{tikzpicture}
    \caption{}\label{fig:smcrun:b}
  \end{subfigure}
  \vrule
  \begin{subfigure}[t]{.32\linewidth}
    \centering
    \begin{tikzpicture}
      \node (pw) at (0,0) [] {\small{\ptid: \store{\xvar}{1}}};
      \node (qw) at ($(pw.west)+(0,-.4)$) [anchor=west] {\small{\pindent\qtid: \store{\xvar}{2}}};
      \node (qu) at ($(qw.west)+(0,-.4)$) [anchor=west] {\small{\pindent\qtid: \textsf{update}}};
      \node (pu) at ($(qu.west)+(0,-.4)$) [anchor=west] {\small{\ptid: \textsf{update}}};
      \node (pr) at ($(pu.west)+(0,-.4)$) [anchor=west] {\small{\ptid: \load{\rreg}{\xvar}}};
    \end{tikzpicture}\\
    \begin{tikzpicture}
      \node (wx1) at (0,0) [] {
        \small{\store{\xvar}{1}}
      };
      \node (rx) at ($(wx1.west)+(1,-.7)$) [anchor=west] {
        \small{\load{\rreg}{\xvar}}
      };
      \node (ux1) at ($(wx1) + (0,-1.5)$) [] {
        \small{\textsf{update}}
      };

      \node (wx2) at (2.1,0) [] {
        \small{\store{\xvar}{2}}
      };
      \node (ux2) at ($(wx2) + (0,-1.5)$) [] {
        \small{\textsf{update}}
      };

      \draw[->,line width=1pt] (wx1) -- (rx);
      \draw[->,line width=1pt] (wx1) -- (ux1);
      \draw[->,line width=1pt] (wx2) -- (ux2);

      \draw[->,line width=1pt] (ux2) -- (ux1);

      \node (p) at ($(wx1)+(0,.5)$) [] {\small{\ptid}};
      \node (q) at ($(wx2)+(0,.5)$) [] {\small{\qtid}};
      \tikzwrapfigbg
    \end{tikzpicture}
    \caption{}\label{fig:smcrun:c}
  \end{subfigure}
  \caption{How SMC with DPOR explores the program of \figurename~\ref{fig:rowe:prog}.}\label{fig:smcrun}
\end{figure}

The algorithm initially explores an arbitrary execution of the program, and
simultaneously generates the corresponding chronological trace. In our
example, this execution can be the one shown in
\figurename~\ref{fig:smcrun:a}, along with its chronological trace.
The algorithm then finds those edges of the chronological trace that can
be reversed by changing the thread scheduling of
the execution. In \figurename~\ref{fig:smcrun:a}, the reversible edges are
the ones from $\ptid: \textsf{update}$ to $\qtid: \textsf{update}$, and
from $\ptid: \load{\rreg}{\xvar}$ to $\qtid: \textsf{update}$.
For each such edge, the program is executed with this edge reversed. Reversing
an edge can potentially lead to a completely different continuation of the
execution, which must then be explored.

In the example, reversing the edge from
$\ptid: \load{\rreg}{\xvar}$ to $\qtid: \textsf{update}$ will generate the
execution and chronological trace in \figurename~\ref{fig:smcrun:b}.
Notice that the new execution is observably different from the
previous one: the load reads the value 2 instead of 1.

The chronological traces in both \figurename~\ref{fig:smcrun:a} and
\ref{fig:smcrun:b} display a reversible edge from
$\ptid: \textsf{update}$ to $\qtid: \textsf{update}$.
The algorithm therefore initiates an execution where $\qtid: \textsf{update}$
is performed before $\ptid: \textsf{update}$.
The algorithm
will generate the
execution and chronological trace in \figurename~\ref{fig:smcrun:c}.

Notice that the only reversible edge in \figurename~\ref{fig:smcrun:c} is
the one from $\qtid: \textsf{update}$ to
$\ptid: \textsf{update}$. However, executing $\ptid: \textsf{update}$ before
$\qtid: \textsf{update}$ has already been explored in \figurename~\ref{fig:smcrun:a} and \figurename~\ref{fig:smcrun:b}. Since there are no more edges that can be reversed, SMC
terminates, having examined precisely the three chronological traces
that exist for the program of \figurename~\ref{fig:rowe:prog}.

\section{Formalization}
\label{sec:formalization}
In this section we summarize our formalization of
the concepts of Section~\ref{sec:intuition}.
We introduce our representation of program executions,
define chronological traces,
formalize Shasha-Snir traces for TSO,
and prove a one-to-one correspondence
between chronological traces and Shasha-Snir traces.
The formalization is
self-contained, but for lack of space, we sometimes use precise English rather
than formal notation. A more fully formalized version can be found in
Appendix~\ref{sec:execs:traces}.

\bjparagraph{Parallel Programs}
We consider parallel programs consisting of
a number of threads that run in parallel,
each executing a deterministic code, written in an assembly-like programming
language. The language includes instructions \store{\xvar}{\rreg},
\load{\rreg}{\xvar}, and \fence.
Other instructions do not access memory, and their precise
syntax and semantics are ignored for brevity.
Here, and in the remainder of this text, \xvar{}, \yvar{}, \zvar{} are
used to name memory locations, \uval{}, \vval{}, \wval{} are used to
name values, and \rreg{}, \sreg{}, \treg{} are used to name processor
registers.
We use \tids{} to denote the set of all thread identifiers.

\bjparagraph{Formal TSO Semantics}
We formalize the TSO model by an operational semantics.
Define a
{\em configuration} as a pair $(\locstate,\mem)$, where
$\mem$ maps memory locations to values, and
$\locstate$ maps each thread $\ptid$ to a local configuration of the form
$\locstate(\ptid) = (\regstate,\bufstate)$, where \regstate{}
is the state of local registers and program counter of $\ptid$, and
\bufstate{} is the
contents of the store buffer of \ptid. This content is a word over pairs
$(\xvar,\vval)$ of memory locations and values.
We let the notation $\bufstate(\xvar)$ denote the value
$\vval$ such that $(\xvar,\vval)$ is the rightmost pair in
\bufstate{} of form $(\xvar,\_)$. If there is no such
pair in \bufstate{}, then $\bufstate(\xvar) = \perp$.

In order to accommodate memory updates in our operational semantics, we
assume that for each thread
$\ptid\in\tids$, there is an auxiliary thread
$\updof{\ptid}$, which
nondeterministically performs memory updates from the store buffer
of \ptid{}.
We use $\auxtids =
\{\updof{\ptid} | \ptid\in\tids\}$ to denote the set of auxiliary
thread identifiers. We use $\ptid$ and $\qtid$ to refer to real
or auxiliary threads in $\tids\cup\auxtids$ as convenient.

For configurations $\aconf = (\locstate,\mem)$ and $\aconf' =
(\locstate',\mem')$, we write
$\aconf\xrightarrow{\ptid}\aconf'$ to denote that from configuration
\aconf{}, thread \ptid{} can execute its next instruction, thereby
changing the configuration into $\aconf'$.
Let $\locstate(\ptid) = (\regstate,\bufstate)$, and
$\regstate_{\pc}$ be obtained from $\regstate$ by advancing the
program counter after $\ptid$ executes its next instruction.
Depending on this next instruction $op$, we have the following cases.

\squeeeeze
\paragraph{Store:}
If $op$ has the form \store{\xvar}{\rreg}, then
$\aconf\xrightarrow{\ptid}\aconf'$ iff $\mem' = \mem$ and $\locstate'
=
\locstate[\ptid\hookleftarrow(\regstate_{\pc},\bufstate\cdot(\xvar,\vval))]$
where $\vval = \regstate(\rreg)$,
i.e., instead of updating the memory,
we insert the entry $(\xvar,\vval)$ at the end of the
store buffer of the thread.

\squeeeeze
\paragraph{Load:}
If $op$ has the form \load{\rreg}{\xvar}, then
$\mem' = \mem$ and either
\begin{enumerate}
\vspace*{-4pt}
\item (\textbf{From memory})
$\bufstate(\xvar) = \perp$ and $\locstate' =
  \locstate[\ptid\hookleftarrow(\regstate_{\pc}[\rreg\hookleftarrow
      \mem(\xvar)],\bufstate)]$,
 i.e., there is no entry for \xvar in the thread's own store
 buffer, so the value is read from memory, or
\item (\textbf{Buffer forwarding})
$\bufstate(\xvar) \neq \perp$ and
  $\locstate' =
  \locstate[\ptid\hookleftarrow(\regstate_{\pc}[\rreg\hookleftarrow\bufstate(\xvar)],\bufstate)]$,
i.e., \ptid reads the value of \xvar from its {\em latest} entry
 in its store buffer.
\end{enumerate}
\squeeeeze
\paragraph{Fence:}
If $op$ has the form \fence, then $\aconf\xrightarrow{\ptid}\aconf'$
iff $\bufstate = \varepsilon$ and $\mem' = \mem$ and $\locstate' =
\locstate[\ptid\hookleftarrow(\regstate_{\pc},\bufstate)]$. A fence
can only be executed when the store buffer of the thread is empty.
\squeeeeze
\paragraph{Update:}
In addition to instructions which are executed by the threads, at any
point when a store buffer is non-empty, an {\em update} event may
nondeterministically occur. The memory is then updated according to
the oldest (leftmost) letter in the store buffer, and that letter is
removed from the buffer.
To formalize this, we will assume that the auxiliary thread
\updof{\ptid} executes a pseudo-instruction $\upd{\xvar}$. We then say
that
$\aconf\xrightarrow{\updof{\ptid}}\aconf'$ iff $\bufstate =
(\xvar,\vval)\cdot\bufstate'$ for some $\xvar$, $\vval$, $\bufstate'$
and $\mem' = \mem[\xvar\hookleftarrow\vval]$ and $\locstate' =
\locstate[\ptid\hookleftarrow(\regstate,\bufstate')]$.

\bjparagraph{Program Executions}
A program execution is a sequence
$\aconf_0\xrightarrow{\ptid_1}\aconf_1\xrightarrow{\ptid_2}\cdots\xrightarrow{\ptid_{n}}\aconf_n$
of configurations related by transitions labelled by actual or
auxiliary thread IDs.
Since each transition of each program thread (including the auxiliary
threads of form $\updof{\qtid}$) is deterministic,
a program run is uniquely determined by its sequence of thread IDs.
We will therefore define an
{\em execution} as a word of {\em events}.
Each event represents a transition in the execution as a triple $(\ptid,i,j)$,
where
$\ptid$  is a regular or auxiliary
thread executing an instruction $i$ (which can possibly be an update), and
the natural number $j$ is such that the event is the $j$th event of
$\ptid$ in the execution.

\bjparagraph{Chronological Traces}
We can now introduce the main conceptual contribution of the paper,
viz.\ {\em chronological traces}.
For an execution $\tau$ we define its chronological trace
$\chronof{\tau}$ as a directed graph $\langle V,E\rangle$.
The vertices $V$ are all the events in $\tau$
(both events representing instructions and events representing updates).
The edges are the union of six relations: $E =
\porel\cup\surel\cup\uurel\cup\csrcrel\cup\ccfrel\cup\ufrel$.
These edge relations are defined as follows,
for two arbitrary events
$\anevent=(\ptid,i,j),\anevent'=(\ptid',i',j')\in V$:

\squeeeze
\paragraph{Program Order:}
$\anevent\porel\anevent'$ iff $\ptid = \ptid'$ and $j' = j+1$,
i.e., $\anevent$ and $\anevent'$ are consecutive events of the same thread.

\squeeeze
\paragraph{Store to Update:}
$\anevent\surel\anevent'$ iff $\anevent'$ is the update event
corresponding to the store $\anevent$.

\squeeeze
\paragraph{Update to Update:}
$\anevent\uurel\anevent'$ iff $i = \upd{\xvar}$ and $i' = \upd{\xvar}$
for some \xvar{}, and $\anevent$ and $\anevent'$ are consecutive updates
to the memory location \xvar{}.

\squeeeze
\paragraph{Source:}
$\anevent\csrcrel\anevent'$ iff $\anevent'$ is a load which reads the
value of the update event $\anevent$, which is from a different process.
Notice that this definition excludes the possibility of $\ptid =
\updof{\ptid'}$; a load is never \textsf{src}-related to an update
from the same thread.

\squeeeze
\paragraph{Conflict:}
$\anevent\ccfrel\anevent'$ iff $\anevent'$ is the update that overwrites
the value read by $\anevent$.

\squeeeze
\paragraph{Update to Fence:}
$\anevent\ufrel\anevent'$ iff $i = \upd{\xvar}$ for some \xvar{}, and
$i' = \fence$ and $\ptid = \updof{\ptid'}$ and $\anevent$ is the
latest update by $\ptid$ which occurs before $\anevent'$ in $\tau$.
The intuition here is that the fence cannot be executed until all
pending updates of the same thread have been flushed from the
buffer. Hence the updates are ordered before the fence, and the chronological
trace has an edge from the last of these updates to the fence event.

\bjparagraph{Shasha-Snir Traces}
We will now formalize Shasha-Snir traces, and
prove that chronological traces are equivalent to
Shasha-Snir traces, in the sense that they induce the same equivalence
relation on executions.
We first recall the definition of Shasha-Snir traces. We follow the
formalization by Bouajjani \etal~\cite{BouajjaniDM13}.

First, we introduce the notion of a completed execution. We
say that an execution $\tau$ is {\em completed} when all stores have
reached memory by means of a corresponding update event.
In the context of Shasha-Snir traces, we will restrict ourselves to completed
executions.

For a completed execution $\tau$, we define the Shasha-Snir trace of $\tau$
as the graph $\tsotraceof{\tau} = \langle V,E\rangle$ where $V$ is the
set of all non-update events $(\ptid,i,j)$ in $\tau$ (i.e., $i \neq
\upd{\xvar}$ for all \xvar{}).
The edges $E$ is the union of four relations
$E=\porel\cup\strel\cup\srcrel\cup\cfrel$, where $\porel$ (program order)
is the same as for Chronological traces, and where, letting
$\anevent=(\ptid,i,j)$ and $\anevent'=(\ptid',i',j')$:

\squeeeeze
\paragraph{Store Order:}
$\anevent\strel\anevent'$ iff $i$ and $i'$ are
two stores, whose corresponding updates are consecutive updates to the
same memory location.
I.e., store order defines a total order on all
the stores to each memory location, based on the order in which they
reach memory.

\squeeeeze
\paragraph{Source:}
$\anevent\srcrel\anevent'$ iff $\anevent'$ is a load which reads its value from
$\anevent$, via memory or by buffer forwarding.

\squeeeeze
\paragraph{Conflict:}
$\anevent\cfrel\anevent'$ iff $\anevent'$ is the store which overwrites the value read by $\anevent$.

We are now ready to state the equivalence theorem.

\begin{theorem}
\label{thm:equivalence}
{\bf (Equivalence of Shasha-Snir traces and chronological traces)}
For a given program $\aprog$ with two completed executions
$\tau,\tau'$, it holds that $\tsotraceof{\tau} = \tsotraceof{\tau'}$
iff $\chronof{\tau} = \chronof{\tau'}$.
\end{theorem}

\noindent
The proof is found in Appendix~\ref{sec:execs:traces}.

\bjparagraph{DPOR for TSO} \label{sect:dpor-tso}
A DPOR algorithm can exploit Chronological traces to perform stateless
model checking of programs that execute under TSO (and PSO), as illustrated at
the end of Section~\ref{sec:intuition}.
The explored executions follow
the semantics of TSO in Section~\ref{sec:formalization}.
For each execution, its happens-before relation,
which is the transitive closure of the edge relation
$E = \porel\cup\surel\cup\uurel\cup\csrcrel\cup\ccfrel\cup\ufrel$
of the corresponding chronological trace, is computed on the fly.
Such a computation is described in more detail in Appendix~\ref{app:algorithm}.
This happens-before relation can in principle be exploited by any DPOR
algorithm to explore at least one execution per equivalence class induced
by Shasha-Snir traces. We state the following theorem
of correctness.

\begin{theorem}
\label{thm:correctness}
{\bf (Correctness of DPOR algorithms)}
The algorithms {\em Source-DPOR} and {\em Optimal-DPOR}
of~\cite{abdulla2014optimal}, based on the
happens-before relation induced by chronological traces, explore at least
one execution per equivalence class induced by Shasha-Snir traces.
Moreover, {\em Optimal-DPOR} explores exactly one execution per equivalence
class.
\end{theorem}
The proof is found in Appendix~\ref{app:algorithm}.

\section{Implementation} \label{sec:implementation}

To show the effectiveness of our techniques we have
implemented a stateless model checker for C programs.
The tool, called Nidhugg, is available as open source at
\url{https://github.com/nidhugg/nidhugg}.
Major design decisions have been that Nidhugg:
\begin{inparaenum}[(i)]
\item should not be bound to a specific hardware architecture and
\item should use an existing, mature implementation of C semantics,
  not implement its own.
\end{inparaenum}
Our choice was to use the LLVM compiler infrastructure~\cite{llvm} and
work at the level of its intermediate representation (IR).
LLVM~IR is low-level and allows us to analyze assembly-like but
target-independent code which is produced after employing all
optimizations and transformations that the LLVM compiler performs till
this stage.

Nidhugg detects
assertion
violations and robustness violations that occur
under the selected memory model.
We implement the Source-DPOR algorithm from Abdulla
\etal~\cite{abdulla2014optimal}, adapted to relaxed memory in the
manner described in this paper.
Before applying Source-DPOR, each spin
loop is replaced by an equivalent single load and assume statement.
This substantially
improves the performance of Source-DPOR, since a waiting spin loop
may generate a huge number of improductive loads, all returning the same
wrong value; all of these loads will cause races, which will cause
the number of explored traces to explode.
Exploration of program executions is performed by interpretation of
LLVM IR, based on the interpreter \textsf{lli} which is distributed
with LLVM.
We support concurrency through the \textsf{pthreads} library. This is
done by hooking calls to pthread functions, and executing changes to
the execution stacks (adding new threads, joining, etc.) as
appropriate within the interpreter.

\section{Experimental Results} \label{sec:experiments}

\newcommand{\bench}[1]{\textsf{\mbox{#1}}}

\newcommand{\bchmkapri}{\textsf{apr\_1.c}\xspace}
\newcommand{\bchmkaprii}{\textsf{apr\_2.c}\xspace}
\newcommand{\bchmkindexer}{\textsf{indexer.c}\xspace}
\newcommand{\bchmkpgsql}{\textsf{pgsql.c}\xspace}
\newcommand{\bchmklamportpsolba}{\textsf{lamport\_fenced\_2\_b2.c}\xspace}
\newcommand{\bchmklamportpsolbb}{\textsf{lamport\_fenced\_2\_b4.c}\xspace}
\newcommand{\bchmklamportpsolbc}{\textsf{lamport\_fenced\_2\_b8.c}\xspace}
\newcommand{\bchmkstacksafe}{\textsf{stack\_safe.c}\xspace}
\newcommand{\bchmkfibtrue}{\textsf{fib\_true.c}\xspace}
\newcommand{\bchmkfibtruejoin}{\bench{fib\_true\_join.c}\xspace}
\newcommand{\bchmkparker}{\bench{parker.c}\xspace}

\newcommand{\SC}{\mbox{SC}\xspace}
\newcommand{\TSO}{\mbox{TSO}\xspace}
\newcommand{\PSO}{\mbox{PSO}\xspace}
\newcommand{\cbmc}{\mbox{CBMC}\xspace}
\newcommand{\gotoinstr}{\mbox{\textsf{goto-instrument}}\xspace}
\newcommand{\dporc}{\mbox{Nidhugg}\xspace}

\newcommand{\loopboundcol}{LB\xspace}

\newcommand{\tabletop}{
\begin{table*}
\centering\footnotesize
\begin{tabular}{l@{\hspace*{0.1em}}c@{\hspace*{0.1em}}c@{\hspace*{0.05em}}r@{\hspace*{0.4em}}r@{\hspace*{0.5em}}r@{\hspace*{0.4em}}r@{\hspace*{0.4em}}r@{\hspace*{0.5em}}r@{\hspace*{0.4em}}r@{\hspace*{0.4em}}r@{\hspace*{0.5em}}r}
\toprule
          & & \multicolumn{1}{c}{} & \multicolumn{3}{c}{\cbmc} & \multicolumn{3}{c}{\gotoinstr} & \multicolumn{3}{c}{\dporc} \\
            \cmidrule(r){4-6}\cmidrule(r){7-9}\cmidrule(r){10-12}
  & fence & \multicolumn{1}{c}{\loopboundcol}  &   \SC &     \TSO &     \multicolumn{1}{r}{\PSO} &   \SC &   \TSO &   \PSO &   \SC &   \TSO &   \PSO \\
\midrule
}
\newcommand{\tablebottom}[2]{
\bottomrule
\end{tabular}
\caption{#1}
\label{#2}
\end{table*}
}

\newcommand{\B}[1]{\textbf{#1}}

\tabletop
apr\_1.c & - & 5 & t/o & t/o & t/o & t/o & ! & ! & \textbf{5.88} & \textbf{6.06} & \textbf{16.98} \\
apr\_2.c & - & 5 & t/o & t/o & t/o & ! & ! & ! & \textbf{2.60} & \textbf{2.20} & \textbf{5.39} \\
dcl\_singleton.c & - & 7 & 5.95 & 31.47 & *18.01 & 5.33 & 5.36 & *0.18 & \textbf{0.08} & \textbf{0.08} & *\textbf{0.08} \\
dcl\_singleton.c & pso & 7 & 5.88 & 30.98 & 29.45 & 5.20 & 5.18 & 5.17 & \textbf{0.08} & \textbf{0.08} & \textbf{0.08} \\
dekker.c & - & 10 & 2.42 & *3.17 & *2.84 & 1.68 & *4.00 & *220.11 & \textbf{0.10} & *\textbf{0.11} & *\textbf{0.09} \\
dekker.c & tso & 10 & 2.39 & 5.65 & *3.51 & 1.62 & 297.62 & t/o & \textbf{0.11} & \textbf{0.12} & *\textbf{0.08} \\
dekker.c & pso & 10 & 2.55 & 5.31 & 4.83 & 1.72 & 428.86 & t/o & \textbf{0.11} & \textbf{0.12} & \textbf{0.12} \\
fib\_false.c & - & - & *1.63 & *3.38 & *3.00 & *\textbf{1.60} & *\textbf{1.58} & *\textbf{1.56} & *2.39 & *5.57 & *6.20 \\
fib\_false\_join.c & - & - & *0.98 & *1.10 & *1.91 & *1.31 & *0.88 & *0.80 & *\textbf{0.32} & *\textbf{0.62} & *\textbf{0.71} \\
fib\_true.c & - & - & \textbf{6.28} & 9.39 & 7.72 & 6.32 & \textbf{7.63} & \textbf{7.62} & 25.83 & 75.06 & 86.32 \\
fib\_true\_join.c & - & - & 6.61 & 8.37 & 10.81 & 7.09 & 5.94 & 5.92 & \textbf{1.20} & \textbf{2.88} & \textbf{3.19} \\
indexer.c & - & 5 & 193.01 & 210.42 & 214.03 & 191.88 & 70.42 & 69.38 & \textbf{0.10} & \textbf{0.09} & \textbf{0.09} \\
lamport.c & - & 8 & 7.78 & *11.63 & *10.53 & 6.89 & t/o & t/o & \textbf{0.08} & *\textbf{0.08} & *\textbf{0.08} \\
lamport.c & tso & 8 & 7.60 & 26.31 & *15.85 & 6.80 & 513.67 & t/o & \textbf{0.09} & \textbf{0.08} & *\textbf{0.07} \\
lamport.c & pso & 8 & 7.72 & 30.92 & 27.51 & 7.43 & t/o & t/o & \textbf{0.08} & \textbf{0.08} & \textbf{0.08} \\
parker.c & - & 10 & 12.34 & *91.99 & *86.10 & 11.63 & \sout{9.70} & \sout{9.65} & \textbf{1.50} & *\textbf{0.09} & *\textbf{0.08} \\
parker.c & pso & 10 & 12.72 & 141.24 & 166.75 & 11.76 & 10.66 & 10.64 & \textbf{1.50} & \textbf{1.92} & \textbf{2.94} \\
peterson.c & - & - & 0.35 & *0.38 & *0.35 & 0.18 & *0.20 & *0.21 & \textbf{0.07} & *\textbf{0.07} & *\textbf{0.07} \\
peterson.c & tso & - & 0.35 & 0.39 & *0.35 & 0.19 & 0.18 & \sout{0.56} & \textbf{0.07} & \textbf{0.07} & *\textbf{0.07} \\
peterson.c & pso & - & 0.35 & 0.41 & 0.40 & 0.18 & 0.18 & 0.19 & \textbf{0.07} & \textbf{0.07} & \textbf{0.08} \\
pgsql.c & - & 8 & 19.80 & 60.66 & *4.63 & 21.03 & 46.57 & *296.77 & \textbf{0.08} & \textbf{0.07} & *\textbf{0.08} \\
pgsql.c & pso & 8 & 23.93 & 71.15 & 121.51 & 19.04 & t/o & t/o & \textbf{0.07} & \textbf{0.07} & \textbf{0.08} \\
pgsql\_bnd.c & pso & (4) & \textbf{3.57} & \textbf{9.55} & \textbf{12.68} & 3.59 & t/o & t/o & 89.44 & 106.04 & 112.60 \\
stack\_safe.c & - & - & 44.53 & 516.01 & 496.36 & 45.11 & 42.39 & 42.50 & \textbf{0.34} & \textbf{0.36} & \textbf{0.43} \\
stack\_unsafe.c & - & - & *1.40 & *1.87 & *2.08 & *1.00 & *0.81 & *0.79 & *\textbf{0.08} & *\textbf{0.08} & *\textbf{0.09} \\
szymanski.c & - & - & 0.40 & *0.44 & *0.43 & 0.23 & *0.89 & *1.16 & \textbf{0.07} & *\textbf{0.13} & *\textbf{0.07} \\
szymanski.c & tso & - & 0.40 & 0.50 & *0.43 & 0.23 & 0.23 & \sout{2.48} & \textbf{0.08} & \textbf{0.08} & *\textbf{0.07} \\
szymanski.c & pso & - & 0.39 & 0.50 & 0.49 & 0.23 & 0.24 & 0.24 & \textbf{0.08} & \textbf{0.08} & \textbf{0.08} \\
\tablebottom{Analysis times (in seconds) for our implementation Nidhugg, as well as CBMC and \textsf{goto-instrument} under the SC, TSO and PSO memory models. Stars (*) indicate that the analysis discovered an error in the benchmark.
A t/o entry means that the tool did not terminate within 10
minutes. An ! entry means that the tool crashed. \sout{Struck-out}
entries mean that the tool gave the wrong result.
In the fence column, a dash (-) means that no fences have been added
to the benchmark, a memory model indicates that fences have been
(manually) added to make the benchmark correct under that and stronger
memory models.
The \loopboundcol column shows the loop unrolling depth. Superior run times are shown in bold face.}{tbl:cmp:dpor:vs:cbmc}

We have applied our implementation to several intensely racy
benchmarks, all implemented in C/pthreads.
They include classical benchmarks, such as Dekker's, Lamport's (fast)
and Peterson's mutual exclusion algorithms.
Other programs, such as \bchmkindexer, are designed to showcase races
that are hard to identify statically.
Yet others (\bchmkstacksafe) use pthread mutexes to entirely avoid
races.
Lamport's algorithm and \bchmkstacksafe originate from the
TACAS Competition on Software Verification (SV-COMP).
Some benchmarks originate from industrial code: \bchmkapri, \bchmkaprii,
\bchmkpgsql and \bchmkparker.

We show the results of our tool \dporc in
Table~\ref{tbl:cmp:dpor:vs:cbmc}. For comparison we also include the
results of two other analysis tools, \cbmc~\cite{AlglaveKT13} and
\gotoinstr~\cite{AlKNT13}, which also target C programs under relaxed
memory. The techniques of \gotoinstr and \cbmc are described in more
detail in Section~\ref{sec:related}.

All experiments were run on
a machine equipped with a 3 GHz Intel i7 processor and 6 GB RAM
running 64-bit Linux.
We use version 4.9 of \gotoinstr and \cbmc.
The benchmarks have been tweaked to work for all tools,
in communication with the developers of \cbmc and \gotoinstr.
All benchmarks are available at
\url{https://github.com/nidhugg/benchmarks_tacas2015}.

Table~\ref{tbl:cmp:dpor:vs:cbmc} shows that our technique performs
well compared to the other tools for most of the examples. We will briefly
highlight a few interesting results.

We see that in most cases Nidhugg pays a very modest performance price
when going from sequential consistency to TSO and PSO.
The explanation is that the number of executions explored by
our stateless model checker is close to the number of Shasha-Snir traces,
which increases very modestly when going from sequential consistency to
TSO and PSO for typical benchmarks.
Consider for example the
benchmark \bchmkstacksafe, which is robust, and therefore has
equally many Shasha-Snir traces (and hence also chronological traces)
under all three memory models. Our technique is able to benefit from
this, and has almost the same run time under TSO and PSO as under SC.

We also see that our implementation compares favourably against \cbmc,
a state-of-the-art bounded model checking tool, and \gotoinstr.
For several benchmarks, our implementation is several orders of magnitude
faster.

The effect of the optimization to replace
each spin loop by a load and assume statement can be seen in the
\textsf{pgsql.c} benchmark.
For comparison, we also include the benchmark
\textsf{pgsql\_bnd.c}, where the spin loop has been modified such that
Nidhugg fails to automatically replace it by an assume statement.

The only other benchmark where Nidhugg is not faster is \bchmkfibtrue.
The benchmark has two threads that perform the actual work, and one
separate thread that checks the correctness of the computed
value, causing many races, as in the case of spin loops.
We show with the benchmark \bchmkfibtruejoin that in this case, the
problem can be alleviated by forcing the threads to join before
checking the result.

Most benchmarks in Table~\ref{tbl:cmp:dpor:vs:cbmc} are small program
cores, ranging from 36 to 118 lines of C code, exhibiting complicated
synchronization patterns.
To show that our technique is also applicable to real life code, we
include the benchmarks \bchmkapri and \bchmkaprii. They each contain
approximately 8000 lines of code taken from the Apache Portable
Runtime library, and exercise the library primitives for thread
management, locking, and memory pools.
Nidhugg is able to analyze the code within a few seconds.
We notice that despite the benchmarks being robust, the analysis under
PSO suffers a slowdown of about three times compared to TSO. This is
because the benchmarks access a large number of different memory
locations. Since PSO semantics require one store buffer per memory
location, this affects analysis under PSO more than under SC and TSO.

\section{Related Work} \label{sec:related}

To the best of our knowledge, our work is the first to apply stateless
model checking techniques to the setting of relaxed memory models; see
e.g.~\cite{abdulla2014optimal} for a recent survey of related work on
stateless model checking and dynamic partial order reduction techniques.
There have been many works dedicated to the verification and checking of programs running under RMM (e.g., \cite{DBLP:journals/jpdc/KrishnamurthyY96,DBLP:journals/tc/LeeP01,LNPVY12,abdulla2012counter,BM08,BurnimSS11,AlglaveM11,BouajjaniDM13,BAM07,YangGLS04}). Some of them propose \emph{precise} analyses for checking safety properties or robustness  of  finite-state programs under TSO (e.g.,~\cite{abdulla2012counter,BouajjaniDM13}). Others describe monitoring and testing techniques for programs under RMM (e.g.,~\cite{BM08,BurnimSS11,LNPVY12}). There are also a number of efforts to design bounded model checking techniques for programs under RMM (e.g.,~\cite{YangGLS04,BAM07}) which encode the verification problem in SAT.

The two closest works to ours are those presented in~\cite{AlglaveKT13,AlKNT13}. The first of them~\cite{AlglaveKT13} develops a bounded model checking technique that can be applied to different memory models (e.g., TSO, PSO, and Power). That technique makes use of the fact that the trace of a program under RMM can be viewed as a partially ordered set. This results in a bounded model checking technique aware of the underlying memory model when constructing the SMT/SAT formula.
The second line of work reduces the verification problem of a program under RMM to verification under SC of a program constructed by a code transformation~\cite{AlKNT13}. This technique tries to encode the effect of the RMM semantics by augmenting the input program with buffers and queues. This work  introduces also the notion of Xtop objects. Although an Xtop object is a valid acyclic representation of Shasha-Snir traces, it will in many cases distinguish executions that are semantically equivalent according to the Shasha-Snir traces. This is never the case for chronological traces. More details on the comparison with Xtop objects can be found in Appendix~\ref{sec:xtop}.
An extensive experimental comparison with the corresponding tools~\cite{AlglaveKT13,AlKNT13} for programs running under the TSO and PSO memory models was given in Section~\ref{sec:experiments}.

\section{Concluding Remarks} \label{sec:conclusion}

We have presented the first technique for efficient \emph{stateless model checking} which is aware of the underlying relaxed memory model. To this end we have introduced \emph{chronological traces} which are novel representations of executions under the TSO and PSO memory models, and induce a happens-before relation that is a partial order and can be used as a basis for DPOR. Furthermore, we have established a strict one-to-one correspondence between chronological and Shasha-Snir traces.
Nidhugg, our publicly available tool, detects bugs in LLVM assembly code produced for C/pthreads programs and can be instantiated to the SC, TSO, and PSO memory models. We have applied Nidhugg to several programs, both benchmarks and of considerable size, and our experimental results show that our technique offers significantly better performance than both CBMC and \textsf{goto-instrument} in many cases.

We plan to extend Nidhugg to more memory models such as Power, ARM, and the C/C++ memory model. This will require to adapt the definition chronological traces to them in order to also guarantee the one-to-one correspondence with Shasha-Snir traces.

\bibliographystyle{abbrv}
\bibliography{bibdatabase,biblio}

\newpage

\appendix

\begin{figure}
  \centering
  \begin{subfigure}[m]{.45\linewidth}
    \centering
    \begin{tikzpicture}
      \node (wx0) at (0,0) [] {
        \small{(\ptid,\st{\xvar},1)}
      };
      \node (es) at ($(wx0.west)+(-.1,0)$) [anchor=east] { $e_s$: };
      \node (wy0) at ($(wx0.west)+(0,-.5)$) [anchor=west] {
        \small{(\ptid,\st{\yvar},2)}
      };
      \node (wz1) at ($(wy0.west)+(0,-.5)$) [anchor=west] {
        \small{\pindent(\qtid,\st{\zvar},1)}
      };
      \node (rx0) at ($(wz1.west)+(0,-.5)$) [anchor=west] {
        \small{(\ptid,\ld{\xvar},3)}
      };
      \node (el) at ($(rx0.west)+(-.1,0)$) [anchor=east] { $e_l$: };
      \node (ux0) at ($(rx0.west)+(0,-.5)$) [anchor=west] {
        \small{(\updof{\ptid},\upd{\xvar},1)}
      };
      \node (eu) at ($(ux0.west)+(-.1,0)$) [anchor=east] { $e_u$: };
      \node (uz1) at ($(ux0.west)+(0,-.5)$) [anchor=west] {
        \small{\pindent(\updof{\qtid},\upd{\zvar},1)}
      };
      \node (uy0) at ($(uz1.west)+(0,-.5)$) [anchor=west] {
        \small{(\updof{\ptid},\upd{\yvar},2)}
      };
    \end{tikzpicture}
    \caption{$\stupdof{e_s} = \roweof{e_l} = e_u$}\label{fig:cmp:examples:rowe:upd}
  \end{subfigure}
  \hspace{3pt}
  \begin{tabular}{@{}c@{}}
  \begin{subfigure}[m]{.5\linewidth}
    \centering
    \begin{tikzpicture}
      \node (wx0) at (0,0) [] {
        \small{(\ptid,\st{\xvar},1)}
      };
      \node (es) at ($(wx0.west)+(-.1,0)$) [anchor=east] { $e_s$: };
      \node (wz1) at ($(wx0.west)+(0,-.5)$) [anchor=west] {
        \small{\pindent(\qtid,\st{\zvar},1)}
      };
      \node (rx0) at ($(wz1.west)+(0,-.5)$) [anchor=west] {
        \small{(\ptid,\ld{\xvar},3)}
      };
      \node (el) at ($(rx0.west)+(-.1,0)$) [anchor=east] { $e_l$: };
      \node (uz1) at ($(rx0.west)+(0,-.5)$) [anchor=west] {
        \small{\pindent(\updof{\qtid},\upd{\zvar},1)}
      };
    \end{tikzpicture}
    \caption{$\stupdof{e_s} = \roweof{e_l} = \evtinf$}\label{fig:cmp:examples:rowe:inf}
  \end{subfigure}\\
  \\
  \begin{subfigure}[m]{.5\linewidth}
    \centering
    \begin{tikzpicture}
      \node (wz0) at (0,0) [] {
        \small{(\ptid,\st{\zvar},1)}
      };
      \node (es) at ($(wz0.west)+(-.1,0)$) [anchor=east] { $e_s$: };
      \node (rx0) at ($(wz0.west)+(0,-.5)$) [anchor=west] {
        \small{(\ptid,\ld{\xvar},3)}
      };
      \node (el) at ($(rx0.west)+(-.1,0)$) [anchor=east] { $e_l$: };
      \node (uz0) at ($(rx0.west)+(0,-.5)$) [anchor=west] {
        \small{(\updof{\ptid},\upd{\zvar},1)}
      };
      \node (eu) at ($(uz0.west)+(-.1,0)$) [anchor=east] { $e_u$: };
    \end{tikzpicture}
    \caption{$\stupdof{e_s} = e_u$ \rule{1pt}{0pt} $\roweof{e_l} = \evtzero$}\label{fig:cmp:examples:zero}
  \end{subfigure}
  \end{tabular}
  \caption{Illustration of the definitions of $\stupdof{}$ and $\roweof{}$.}\label{fig:cmp:examples}
\end{figure}

\begin{figure*}[b]
  \begin{tabular}[b]{@{}cc@{}}
      \begin{subfigure}[m]{.36\linewidth}
        \centering
        \small{
          \begin{tabular}{@{}l@{\hspace{1em}}l@{}}
            \multicolumn{1}{c}{\ptid{}} & \multicolumn{1}{c}{\qtid{}} \\
            \store{\xvar}{1} & \store{\yvar}{1} \\
            \store{\zvar}{1} & \store{\zvar}{0} \\
            \load{\rreg}{\zvar} & \load{\rreg}{\zvar}\\
            \load{\sreg}{\yvar} & \load{\sreg}{\xvar}\\
          \end{tabular}
        }
        \caption{A small program illustrating the idiom of Peterson's
          mutual exclusion algorithm.}\label{fig:peterson:prog}
      \end{subfigure}
      &
      \begin{subfigure}[m]{.59\linewidth}
      \centering
      \begin{tabular}{@{}c@{}}
      \begin{tikzpicture}
        \node (wx0) at (0,0) [] {
          \small{\st{\xvar}}
        };
        \node (wz0) at ($(wx0.west)+(0,-1.5)$) [anchor=west] {
          \small{\st{\zvar}}
        };
        \node (rz0) at ($(wz0.west)+(0,-1.5)$) [anchor=west] {
          \small{\ld{\zvar}}
        };
        \node (ry0) at ($(rz0.west)+(0,-1.5)$) [anchor=west] {
          \small{\ld{\yvar}}
        };
        \node (ux0) at ($(wx0.west)+(2,-.75)$) [anchor=west] {
          \small{\upd{\xvar}}
        };
        \node (uz0) at ($(ux0.west)+(0,-1.5)$) [anchor=west] {
          \small{\upd{\zvar}}
        };
        \node (wy1) at (6,0) [] {
          \small{\st{\yvar}}
        };
        \node (wz1) at ($(wy1.west)+(0,-1.5)$) [anchor=west] {
          \small{\st{\zvar}}
        };
        \node (rz1) at ($(wz1.west)+(0,-1.5)$) [anchor=west] {
          \small{\ld{\zvar}}
        };
        \node (rx1) at ($(rz1.west)+(0,-1.5)$) [anchor=west] {
          \small{\ld{\xvar}}
        };
        \node (uy1) at ($(wy1.west)+(-2,-.75)$) [anchor=west] {
          \small{\upd{\yvar}}
        };
        \node (uz1) at ($(uy1.west)+(0,-1.5)$) [anchor=west] {
          \small{\upd{\zvar}}
        };

        \node (q) at ($(wx0.north) + (0,.5)$) [] {\ptid\vphantom{d}};
        \node (qupd) at ($(ux0.north) + (0,1.25)$) [] {\updof{\ptid}};
        \node (p) at ($(wy1.north) + (0,.5)$) [] {\qtid\vphantom{d}};
        \node (pupd) at ($(uy1.north) + (0,1.25)$) [] {\updof{\qtid}};

        \draw[->,line width=1pt] (wx0) -- node [left] {\small{\textsf{po}}} (wz0);
        \draw[->,line width=1pt] (wz0) -- node [left] {\small{\textsf{po}}} (rz0);
        \draw[->,line width=1pt] (rz0) -- node [left] {\small{\textsf{po}}} (ry0);
        \draw[->,line width=1pt] (wx0) -- node [above] {\small{\textsf{su}}} (ux0);
        \draw[->,line width=1pt] (wz0) -- node [above] {\small{\textsf{su}}} (uz0);
        \draw[->,line width=1pt] (ux0) -- node [left] {\small{\textsf{po}}} (uz0);

        \draw[->,line width=1pt] (wy1) -- node [right] {\small{\textsf{po}}} (wz1);
        \draw[->,line width=1pt] (wz1) -- node [right] {\small{\textsf{po}}} (rz1);
        \draw[->,line width=1pt] (rz1) -- node [right] {\small{\textsf{po}}} (rx1);
        \draw[->,line width=1pt] (wy1) -- node [above] {\small{\textsf{su}}} (uy1);
        \draw[->,line width=1pt] (wz1) -- node [above] {\small{\textsf{su}}} (uz1);
        \draw[->,line width=1pt] (uy1) -- node [right] {\small{\textsf{po}}} (uz1);

        \draw[->,line width=1pt] (uz0) -- node [below] {\small{\textsf{uu}}} (uz1);
        \draw[->,line width=1pt,out=325,in=180] (ux0) to node [pos=.7,right] {\small{\textsf{src}}} (rx1);
        \draw[->,line width=1pt,out=0,in=210] (ry0) to node [left] {\small{\textsf{cf}}} (uy1);
        \draw[->,line width=1pt,out=325,in=270] (rz0) to node [below] {\small{\textsf{cf}}} (uz1);
        \tikzwrapfigbg
      \end{tikzpicture}\\
      \end{tabular}
      \caption{The chronological trace $\chronof{\tau}$ corresponding to the execution in Figure~\ref{fig:peterson:cmp}. Notice that there is no edge between $(\qtid,\ld{\zvar},3)$ and either of the updates to \zvar.}\label{fig:peterson:chronotrace}
      \end{subfigure}
      \\
      \begin{subfigure}[m]{.38\linewidth}
        \centering
        \small{
          \begin{tabular}{l}
            (\ptid, \st{\xvar},1) \\
            \pindent(\qtid, \st{\yvar},1) \\
            (\ptid, \st{\zvar},2) \\
            (\ptid, \ld{\zvar},3) \\
            (\ptid, \ld{\yvar},4) \\
            \pindent(\updof{\qtid}, \upd{\yvar},1) \\
            \pindent(\qtid, \st{\zvar},2) \\
            \pindent(\qtid, \ld{\zvar},3) \\
            (\updof{\ptid}, \upd{\xvar},1) \\
            \pindent(\qtid, \ld{\xvar},4) \\
            (\updof{\ptid}, \upd{\zvar},2) \\
            \pindent(\updof{\qtid}, \upd{\zvar},2) \\
          \end{tabular}
        }
        \caption{An execution $\tau$.}\label{fig:peterson:cmp}
      \end{subfigure}
    &
    \begin{subfigure}[m]{.57\linewidth}
      \centering
      \begin{tabular}{@{}c@{}}
      \begin{tikzpicture}
        \node (wx0) at (0,0) [] {
          \small{\st{\xvar}}
        };
        \node (wz0) at ($(wx0.west)+(0,-1.5)$) [anchor=west] {
          \small{\st{\zvar}}
        };
        \node (rz0) at ($(wz0.west)+(0,-1.5)$) [anchor=west] {
          \small{\ld{\zvar}}
        };
        \node (ry0) at ($(rz0.west)+(0,-1.5)$) [anchor=west] {
          \small{\ld{\yvar}}
        };
        \node (wy1) at (4,0) [] {
          \small{\st{\yvar}}
        };
        \node (wz1) at ($(wy1.west)+(0,-1.5)$) [anchor=west] {
          \small{\st{\zvar}}
        };
        \node (rz1) at ($(wz1.west)+(0,-1.5)$) [anchor=west] {
          \small{\ld{\zvar}}
        };
        \node (rx1) at ($(rz1.west)+(0,-1.5)$) [anchor=west] {
          \small{\ld{\xvar}}
        };

        \node (q) at ($(wx0.north) + (0,.5)$) [] {\ptid\vphantom{d}};
        \node (p) at ($(wy1.north) + (0,.5)$) [] {\qtid\vphantom{d}};

        \draw[->,line width=1pt] (wx0) -- node [left] {\small{\textsf{po}}} (wz0);
        \draw[->,line width=1pt,out=260,in=100] (wz0) to node [left] {\small{\textsf{po}}} (rz0);
        \draw[->,line width=1pt,out=280,in=80] (wz0) to node [right] {\small{\textsf{src}}} (rz0);
        \draw[->,line width=1pt] (rz0) -- node [left] {\small{\textsf{po}}} (ry0);

        \draw[->,line width=1pt] (wy1) -- node [right] {\small{\textsf{po}}} (wz1);
        \draw[->,line width=1pt,out=280,in=80] (wz1) to node [right] {\small{\textsf{po}}} (rz1);
        \draw[->,line width=1pt,out=260,in=100] (wz1) to node [left] {\small{\textsf{src}}} (rz1);
        \draw[->,line width=1pt] (rz1) -- node [right] {\small{\textsf{po}}} (rx1);

        \draw[->,line width=1pt] (wz0) -- node [above] {\small{\textsf{st}}} (wz1);
        \draw[->,line width=1pt,out=325,in=180] (wx0) to node [right,pos=.7] {\small{\textsf{src}}} (rx1);
        \draw[->,line width=1pt,out=0,in=210] (ry0) to node [above=10pt,pos=.8] {\small{\textsf{cf}}} (wy1);
        \draw[->,line width=1pt] (rz0) -- node [above] {\small{\textsf{cf}}} (wz1);
        \tikzwrapfigbg
      \end{tikzpicture}\\
      \end{tabular}
      \caption{The Shasha-Snir trace $\tsotraceof{\tau}$ corresponding to the
        execution in
        Figure~\ref{fig:peterson:cmp}.}\label{fig:peterson:tsotrace}
    \end{subfigure}\\
  \end{tabular}
  \caption{Traces illustrated by the idiom of Peterson's
    mutual exclusion algorithm.}\label{fig:peterson}
\end{figure*}
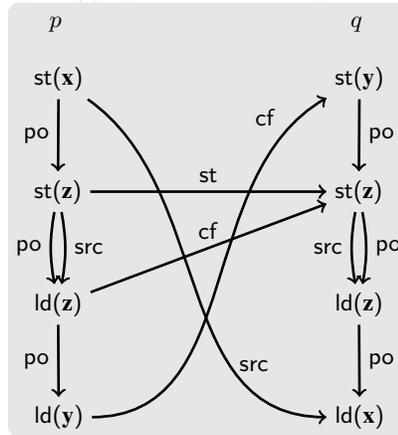

\section{Executions and Traces}\label{sec:execs:traces}

In this appendix we introduce our representation of program executions
and define chronological traces.
We also formalize Shasha-Snir traces for TSO
and prove that there is a one-to-one correspondence
between our chronological traces and Shasha-Snir traces. For completeness,
we first recall, in slightly more detail, our model of TSO and
our definition of executions.

\subsection{Concurrent Programs}\label{sec:concurrent}

We consider parallel programs running on shared memory under TSO.
We assume that a number of threads run in parallel, each executing a
deterministic code. We assume an assembly-like programming language
for the code. The language includes instructions \store{\xvar}{\rreg},
\load{\rreg}{\xvar}, and \fence.
Other instructions do not access memory, and their precise
syntax and semantics are ignored in this text for brevity.
Here, and in the remainder of this text, \xvar{}, \yvar{}, \zvar{} are
used to name memory locations, \uval{}, \vval{}, \wval{} are used to
name values, and \rreg{}, \sreg{}, \treg{} are used to name registers
local to processor cores.
Where convenient, we will use the short forms \st{\xvar} and
\ld{\xvar} to denote some store and load of \xvar{} respectively,
where the value is not interesting.
We will use \tids{} to denote the set of all thread identifiers, and
\memlocs{} to denote the set of all memory locations.

\subsection{TSO Semantics}\label{sec:semantics}

We now define the semantics of our programming
language when running under TSO.
For a function $f$, we use the notation $f[x\hookleftarrow v]$ to
denote the function $f'$ such that $f'(x) = v$ and $f'(y) = f(y)$
whenever $y \neq x$.
We use $w\cdot{}w'$ to denote the concatenation of the words $w$ and $w'$.

Let the program \aprog{} be given. We define a system
configuration as a pair $(\locstate,\mem)$, where 
$\locstate(\ptid)$ defines the thread local configuration for each
thread $\ptid$, and $\mem(\xvar)$ is the current value at
\xvar{} in memory. The local configuration of each thread is defined
by $\locstate(\ptid) = (\regstate,\bufstate)$. Here \regstate{}
is some structure which contains local information such as current
register valuation (denoted $\regstate(\rreg)$) and current program
counter (denoted $\regstate(\pc)$). \bufstate{} represents the
contents of the store buffer of the thread \ptid. It is a word over pairs
$(\xvar,\vval)$ of memory locations and values.
We let the notation $\bufstate(\xvar)$ denote the value
$\vval$ such that $(\xvar,\vval)$ is the rightmost letter in
\bufstate{} which is of the form $(\xvar,\_)$. If there is no such
letter in \bufstate{}, then we say that $\bufstate(\xvar) = \perp$.

In order to accommodate memory updates in our operational semantics we
will introduce the notion of {\em auxiliary threads}. For each thread
$\ptid\in\tids$, we assume that there is an auxiliary thread
$\updof{\ptid}$. The auxiliary thread $\updof{\ptid}$ will
nondeterministically perform memory updates from the store buffer
of \ptid{}, when the buffer is non-empty. We use $\auxtids =
\{\updof{\ptid} | \ptid\in\tids\}$ to denote the set of auxiliary
thread identifiers. We will use $\ptid$ and $\qtid$ to refer to real
or auxiliary threads in $\tids\cup\auxtids$ as convenient.

For configurations $\aconf = (\locstate,\mem)$ and $\aconf' =
(\locstate',\mem')$, we use the notation
$\aconf\xrightarrow{\ptid}\aconf'$ to denote that from configuration
\aconf{}, thread \ptid{} can execute its next instruction, and doing
so will change the system configuration into $\aconf'$. We define the
transition relation $\aconf\xrightarrow{\ptid}\aconf'$ depending on
what the next instruction $op$ of \ptid{} is in $\aconf$. In the
following we assume $\aconf = (\locstate,\mem)$ and $\aconf' =
(\locstate',\mem')$ and $\locstate(\ptid) =
(\regstate,\bufstate)$.
We let $\regstate_{\pc} = \regstate[\pc\hookleftarrow pc']$ where
$pc'$ is the next program counter of $\ptid$ after executing $op$.

\paragraph{Store:}
If $op$ has the form \store{\xvar}{\rreg}, then
$\aconf\xrightarrow{\ptid}\aconf'$ iff $\mem' = \mem$ and $\locstate'
=
\locstate[\ptid\hookleftarrow(\regstate_{\pc},\bufstate\cdot(\xvar,\vval))]$
where $\vval = \regstate(\rreg)$.
Intuitively, under TSO, instead of updating the memory with the new
value \vval, we insert the entry $(\xvar,\vval)$ at the end of the
store buffer of the thread.

\paragraph{Load:}
If $op$ has the form \load{\rreg}{\xvar}, then
$\aconf\xrightarrow{\ptid}\aconf'$ iff either
\begin{enumerate}
\item (\textbf{From memory})
$\bufstate(\xvar) = \perp$ and $\mem' = \mem$ and $\locstate' =
  \locstate[\ptid\hookleftarrow(\regstate_{\pc}[\rreg\hookleftarrow
      \mem(\xvar)],\bufstate)]$.
 Intuitively, there is no entry for \xvar in the thread's own store
 buffer, so the value is read from memory.
\item (\textbf{Buffer forwarding})
$\bufstate(\xvar) \neq \perp$ and $\mem' = \mem$ and
  $\locstate' =
  \locstate[\ptid\hookleftarrow(\regstate_{\pc}[\rreg\hookleftarrow\bufstate(\xvar)],\bufstate)]$.
 Intuitively, we read the value of \xvar from its {\em latest} entry
 in the store buffer of the thread.
\end{enumerate}

\paragraph{Fence:}
If $op$ has the form \fence, then $\aconf\xrightarrow{\ptid}\aconf'$
iff $\bufstate = \varepsilon$ and $\mem' = \mem$ and $\locstate' =
\locstate[\ptid\hookleftarrow(\regstate_{\pc},\bufstate)]$. A fence
can be only executed when the store buffer of the thread is empty.

\paragraph{Update:}
In addition to instructions which are executed by the threads, at any
point when a store buffer is non-empty, an {\em update} event may
nondeterministically occur. The memory is then updated according to
the oldest (leftmost) letter in the store buffer, and that letter is
removed from the buffer.
To formalize this, we will assume that the auxiliary thread
\updof{\ptid} executes a pseudo-instruction $\upd{\xvar}$. We then say
that
$\aconf\xrightarrow{\updof{\ptid}}\aconf'$ iff $\bufstate =
(\xvar,\vval)\cdot\bufstate'$ for some $\xvar$, $\vval$, $\bufstate'$
and $\mem' = \mem[\xvar\hookleftarrow\vval]$ and $\locstate' =
\locstate[\ptid\hookleftarrow(\regstate,\bufstate')]$.

\subsection{Program Executions}
\label{sec:executions}
Based on the operational semantics in Section~\ref{sec:semantics}, a program
execution can be defined as a sequence
$\aconf_0\xrightarrow{\ptid_0}\aconf_1\xrightarrow{\ptid_1}\cdots\xrightarrow{\ptid_{n-1}}\aconf_n$
of configurations related by transitions labelled by actual or
auxiliary thread IDs.
Since each transition of each program thread (including the auxiliary
threads of form $\updof{\qtid}$) is deterministic,
a program run is uniquely determined by its sequence of thread IDs. 
Formally, we will therefore define each
{\em execution} as a word of {\em events}.
Each event is a triple $(\ptid,i,j)$
which represents one transition in
the run.
Here the thread $\ptid\in\tids\cup\auxtids$ can be either a regular
thread $\ptid\in\tids$ executing an instruction $i$, or an auxiliary
thread $\ptid\in\auxtids$ performing an update to memory from a store
buffer. In the latter case, $i = \upd{\xvar}$ is the instruction that
denotes an update to a memory location.
The natural number $j$ is used to disambiguate events. We
let $j$ equal one plus the number of preceding triples
$(\ptid',i',j')$ in the execution with $\ptid' = \ptid$. For an
event $\anevent = (\ptid,i,j)$, we define $\tidof{\anevent} = \ptid$.
We will use $\events$ to denote the set of all possible events.
Figure~\ref{fig:cmp:examples} shows three sample executions.

For an execution $\tau$ and two events $\anevent,\anevent'$ in $\tau$,
we say that $\anevent\cmplt{\tau}\anevent'$ iff $\anevent$ strictly
precedes $\anevent'$ in $\tau$.
We define two dummy events $\evtzero = (\perp,\perp,-1)$ and $\evtinf
= (\perp,\perp,\infty)$, and we extend $\cmplt{\tau}$ such that for
every event $\anevent \not\in \{\evtzero,\evtinf\}$ we have $\evtzero
\cmplt{\tau} \anevent \cmplt{\tau} \evtinf$.

For an execution $\tau$ and an event $\anevent =
(\ptid,\st{\xvar},j)$ in $\tau$, we define $\stupdof{\anevent}$ to be
the update event in $\tau$ corresponding to the store event
$\anevent$. Formally, let $k$ be the number of events $\anevent_w =
(\ptid',\st{\yvar},j')$ for any memory location \yvar{} in $\tau$ such
that $\ptid' = \ptid$ and $j' \leq j$. Then $\stupdof{\anevent} =
(\updof{\ptid},\upd{\xvar},k)$ if there is such an event in~$\tau$.
Otherwise $\stupdof{\anevent} = \evtinf$, denoting that the update is
still pending at the end of $\tau$.
Figure~\ref{fig:cmp:examples:rowe:upd} illustrates the typical case,
where the store $\anevent_s$ is eventually followed by its
corresponding update $\stupdof{\anevent_s} = \anevent_u$.
Figure~\ref{fig:cmp:examples:rowe:inf} shows the case when
the update corresponding to the store $\anevent_s$ is still pending at
the end of the execution, and therefore $\stupdof{\anevent_s} = \evtinf$.

For an execution $\tau$ and an event $\anevent = (\ptid,\ld{\xvar},j)$
in $\tau$, we define $\roweof{\anevent}$ to be the update event of the
latest store to \xvar, which precedes $\anevent$ in the same thread.
The intuition is that $\roweof{\anevent}$ is the update from which
$\anevent$ would get its value in the case of buffer forwarding.
Formally, if there is an event $\anevent_w = (\ptid,\st{\xvar},k)$ in
$\tau$ such that $k < j$ and there is no event $(\ptid,\st{\xvar},l)$
in $\tau$ with $k < l < j$, then $\roweof{\anevent} =
\stupdof{\anevent_w}$. Otherwise $\roweof{\anevent} = \evtzero$.
Figures~\ref{fig:cmp:examples:rowe:upd} and
\ref{fig:cmp:examples:rowe:inf} show the typical case, where
$\roweof{\anevent_l}$ is taken to be the update
corresponding to the latest preceding store by the same thread to the
same memory location. Figure~\ref{fig:cmp:examples:zero} shows the
case when there is no such preceding store, and $\roweof{\anevent_l}$
is taken to be the dummy event $\evtzero$. (Notice that the store
$\anevent_s$ is to a different memory location.)

\subsection{Chronological Traces}
\label{sec:traces}

We can now introduce the main conceptual contribution of the paper,
viz.\ {\em chronological traces}.
For an execution $\tau$ we define its chronological trace
$\chronof{\tau}$ as a directed graph $\langle V,E\rangle$.
The vertices $V$ are all the events in $\tau$; both events
representing instructions and events representing updates.
The edges are the union of six relations: $E =
\porel\cup\surel\cup\uurel\cup\csrcrel\cup\ccfrel\cup\ufrel$.

We will illustrate the definition on an execution of the program in
Figure~\ref{fig:peterson:prog}, which contains an
idiom that occurs in the mutual exclusion algorithm of Peterson. It is
mostly the same as that from Dekker's mutual exclusion algorithm. But
it has two additional accesses in each thread to a separate memory
location \zvar. These provide an opportunity to display buffer
forwarding. Figure~\ref{fig:peterson:cmp} shows an example of an
execution and Figure~\ref{fig:peterson:chronotrace} shows
its corresponding chronological trace.

We define the edge relations of chronological traces as follows,
for two arbitrary events
$\anevent=(\ptid,i,j),\anevent'=(\ptid',i',j')\in V$:

\paragraph{Program Order:}
$\anevent\porel\anevent'$ iff $\ptid = \ptid'$ and $j' = j+1$.
For example, in Figure~\ref{fig:peterson:chronotrace} there is a
program order edge from the store instruction $(\ptid,\st{x},1)$ to
the store instruction $(\ptid,\st{z},2)$ which immediately follows it
in the program of thread $\ptid$. Similarly, the updates of each
thread are program ordered. E.g.,
$(\updof{\ptid},\upd{\xvar},1)\porel(\updof{\ptid},\upd{\zvar},2)$.

\paragraph{Store to Update:}
$\anevent\surel\anevent'$ iff $i = \st{\xvar}$ for some $\xvar$ and
$\stupdof{\anevent} = \anevent'$. I.e., $\anevent'$ is the update
corresponding to the store $\anevent$.
This is illustrated in Figure~\ref{fig:peterson:chronotrace} where
there is an \textsf{su}-edge from each store, to its corresponding
update.

\paragraph{Update to Update:}
$\anevent\uurel\anevent'$ iff $i = \upd{\xvar}$ and $i' = \upd{\xvar}$
for some \xvar{} and $\anevent\cmplt{\tau}\anevent'$ and there is no
event $\anevent'' = (\ptid'',\upd{\xvar},j'')$ such that
$\anevent\cmplt{\tau}\anevent''\cmplt{\tau}\anevent'$. I.e., $\uurel$
defines the total, chronological order on updates to each memory
location.
In Figure~\ref{fig:peterson:chronotrace} we see that the two updates
to \zvar{} are \textsf{uu}-ordered with each other in the same order
as they appear in the execution in
Figure~\ref{fig:peterson:cmp}. However, they are not \textsf{uu}-ordered
with the updates to \xvar{} and \yvar{}.

\paragraph{Source:}
$\anevent\csrcrel\anevent'$ iff for some \xvar{} it
holds that $i = \upd{\xvar}$ and $i' = \ld{\xvar}$ and
$\roweof{\anevent'}\cmplt{\tau}\anevent\cmplt{\tau}\anevent'$
and there is no update $\anevent'' = (\ptid'',\upd{\xvar},j'')$ to
\xvar such that
$\anevent\cmplt{\tau}\anevent''\cmplt{\tau}\anevent'$.
I.e., if the source of the value read by $\anevent'$ is an update
$\anevent$ from a different process, then
$\anevent\csrcrel\anevent'$. Otherwise, there is no incoming
$\csrcrel$ edge to $\anevent'$.
Notice that this definition excludes the possibility of $\ptid =
\updof{\ptid'}$; a load is never \textsf{src}-related to an update
from the same thread.
In Figure~\ref{fig:peterson:chronotrace} we see that the load
$(\qtid,\ld{\xvar},4)$ takes its value from the update
$(\updof{\ptid},\upd{\xvar},1)$. Therefore the events are
\textsf{src}-related. But the loads to \zvar{} both read the value
written by their own thread, and therefore have no \textsf{src}-relation.

\paragraph{Conflict:}
$\anevent\ccfrel\anevent'$ iff $i = \ld{\xvar}$ and $i' = \upd{\xvar}$
for some \xvar{} and $\anevent'$ is the first (w.r.t. $\cmplt{\tau}$)
event $\anevent_u$ of the form $(\_,\upd{\xvar},\_)$ such that both
$\anevent\cmplt{\tau}\anevent_u$ and
$\roweof{\anevent}\cmplt{\tau}\anevent_u$.
The intuition here is that $\anevent\ccfrel\anevent'$ when $\anevent'$
is the first update which succeeds $\anevent$ in the coherence order
of $\xvar$. Equivalently, $\anevent'$ is the update that overwrites
the value that was read by $\anevent$.
In Figure~\ref{fig:peterson:chronotrace}, the load to \yvar{} by
\ptid{} reads the initial value of \yvar{}, which is then overwritten
by the update to \yvar{} by \qtid{}.
Therefore the load has a \textsf{cf}-edge to the update.
The load to \zvar{} by \ptid{} reads the value of
$(\updof{\ptid},\upd{\zvar},2)$ by buffer forwarding. That value is
later overwritten in memory by the update
$(\updof{\qtid},\upd{\zvar},2)$. Therefore the load has a
\textsf{cf}-edge to the update originating in thread $\qtid$.

\paragraph{Update to Fence:}
$\anevent\ufrel\anevent'$ iff $i = \upd{\xvar}$ for some \xvar{}, and
$i' = \fence$ and $\ptid = \updof{\ptid'}$ and
$\anevent\cmplt{\tau}\anevent'$ and there is no event $\anevent'' =
(\ptid,\upd{\yvar},j'')$ for any \yvar{} such that $\anevent
\cmplt{\tau}\anevent''\cmplt{\tau}\anevent'$.
The intuition here is that the fence cannot be executed until all
pending updates of the same thread have been flushed from the
buffer. Hence the updates are ordered before the fence.

\subsection{Shasha-Snir Traces}
\label{sec:sstraces}
We will now prove that chronological traces are equivalent to
Shasha-Snir traces, in the sense that
there is a one-to-one mapping between Shasha-Snir traces $T$ and chronological
traces $T_C$ such that the set of executions corresponding to $T$ is
the same as the set of executions corresponding to $T_C$.

We briefly recall the definition of Shasha-Snir traces, based on the
definition by Bouajjani \etal~\cite{BouajjaniDM13}.

First, we introduce the notion of a {\em completed} execution. We
say that an execution $\tau$ is completed when all stores have
reached memory, i.e., when for every event $\anevent =
(\ptid,\st{\xvar},j)$ in $\tau$ we have $\stupdof{\anevent} \neq
\evtinf$.
In the context of Shasha-Snir traces, we will restrict ourselves to completed
executions.

For a completed execution $\tau$, we define the Shasha-Snir trace of $\tau$
as the graph $\tsotraceof{\tau} = \langle V,E\rangle$ where $V$ is the
set of all non-update events $(\ptid,i,j)$ in $\tau$ where $i \neq
\upd{\xvar}$ for all \xvar{}.
The edges $E$ is the union of four relations
$E=\porel\cup\strel\cup\srcrel\cup\cfrel$.

For two arbitrary events
$\anevent=(\ptid,i,j),\anevent'=(\ptid',i',j')\in V$, we define the
relations as follows:

\paragraph{Program Order:}
$\anevent\porel\anevent'$ iff $\ptid = \ptid'$ and $j' = j+1$. This is
the same program order as for chronological traces.

\paragraph{Store Order:}
$\anevent\strel\anevent'$ iff $i = \st{\xvar}$ and $i' = \st{\xvar}$
and the corresponding updates are ordered in $\tau$
s.t. $\stupdof{\anevent} \cmplt{\tau} \stupdof{\anevent'}$ and there
is no other update event $\anevent'' = (\ptid'',\upd{\xvar},j'')$ such
that $\stupdof{\anevent} \cmplt{\tau} \anevent'' \cmplt{\tau}
\stupdof{\anevent'}$. I.e., store order defines a total order on all
the stores to each memory location, based on the order in which they
reach memory.

\paragraph{Source:}
$\anevent\srcrel\anevent'$ iff $i' = \ld{\xvar}$ and $\anevent$ is the
maximal store event $\anevent'' = (\ptid'',\st{\xvar},j'')$ with respect to
$\strel^*$ such that either $\stupdof{\anevent''}\cmplt{\tau}\anevent'$ or
$\anevent''\porel^*\anevent'$. I.e., $\anevent\srcrel\anevent'$
when $\anevent'$ is a load which reads its value from
$\anevent$, via memory or by buffer forwarding.

\paragraph{Conflict:}
$\anevent\cfrel\anevent'$ iff $i = \ld{\xvar}$ and $i' = \st{\xvar}$
and if there is an event $\anevent''$ such that
$\anevent''\srcrel\anevent$ then $\anevent''\strel\anevent'$,
otherwise $\anevent'$ has no predecessor in $\strel$.
I.e., $\anevent'$ is the store which overwrites the value that was
read by $\anevent$.

\paragraph{}
The definition of Shasha-Snir traces is illustrated in
Figure~\ref{fig:peterson:tsotrace}.
We are now ready to state the equivalence theorem.

\addtocounter{theorem}{-2}
\begin{theorem}
{\bf (Equivalence of Shasha-Snir traces and chronological traces)}
For a given program $\aprog$ with two completed executions
$\tau,\tau'$, it holds that $\tsotraceof{\tau} = \tsotraceof{\tau'}$
iff $\chronof{\tau} = \chronof{\tau'}$.
\end{theorem}

\paragraph{Proof}
We decompose the theorem into the following two lemmas, which are
proven separately.

\begin{lemma}\label{lemma:equiv:ss:chrono:right}
{\bf (Equivalence of Shasha-Snir traces and chronological traces: $\Rightarrow$ direction)}
For a given program $\aprog$ with two completed executions
$\tau,\tau'$, it holds that if $\tsotraceof{\tau} = \tsotraceof{\tau'}$
then $\chronof{\tau} = \chronof{\tau'}$.
\end{lemma}

\begin{lemma}\label{lemma:equiv:ss:chrono:left}
{\bf (Equivalence of Shasha-Snir traces and chronological traces: $\Leftarrow$ direction)}
For a given program $\aprog$ with two completed executions
$\tau,\tau'$, it holds that if $\chronof{\tau} = \chronof{\tau'}$ then
$\tsotraceof{\tau} = \tsotraceof{\tau'}$.
\end{lemma}

\paragraph{Proof of Lemma~\ref{lemma:equiv:ss:chrono:right}}
Let two completed executions $\tau$ and $\tau'$ be given. Let\\
$\tsotraceof{\tau} = \langle
V_{SS},\porelof{\tau}\cup\strelof{\tau}\cup\srcrelof{\tau}\cup\cfrelof{\tau}\rangle$ and\\
$\tsotraceof{\tau'} = \langle V_{SS}',
\porelof{\tau'}\cup\strelof{\tau'}\cup\srcrelof{\tau'}\cup\cfrelof{\tau'}\rangle$ and\\
$\chronof{\tau} = \langle V_C,\porelof{\tau}\cup\surelof{\tau}\cup\uurelof{\tau}\cup\csrcrelof{\tau}\cup\ccfrelof{\tau}\cup\ufrelof{\tau}\rangle$ and \\
$\chronof{\tau'} = \langle V'_C,\porelof{\tau'}\cup\surelof{\tau'}\cup\uurelof{\tau'}\cup\csrcrelof{\tau'}\cup\ccfrelof{\tau'}\cup\ufrelof{\tau'}\rangle$.\\
Furthermore, assume that $\tsotraceof{\tau} = \tsotraceof{\tau'}$.

First, we determine that the events are the same in both chronological
traces: $V_C = V'_C$. From $V_{SS} = V'_{SS}$ we have that the
non-update events in $\tau$ are the same as the ones in $\tau'$. Since
$\tau$ and $\tau'$ contain the same stores for each thread in the same
per-thread order, it follows from the completedness of $\tau$ and
$\tau'$, and from the TSO semantics that $\tau$ and $\tau'$ also have
the same update events. Hence $V_C = V'_C$.

We see that the definitions of program order and store to update order
in chronological traces are entirely determined by which events exist
in the execution for each thread. Since both executions have the same
events, we conclude that $\porelof{\tau} = \porelof{\tau'}$ and $\surelof{\tau} =
\surelof{\tau'}$. The equality of update to fence order follows similarly.

Let us consider the definitions of update to update order for
chronological traces and store order for Shasha-Snir traces. We see
that there is a one-to-one mapping between relations
$\anevent\strelof{\tau}\anevent'$ for stores in Shasha-Snir traces to
relations $\stupdof{\anevent}\uurelof{\tau}\stupdof{\anevent'}$ in
chronological traces. Since the store orders are the same for $\tau$
and $\tau'$, we thus conclude that the update to update orders are
also the same: $\uurelof{\tau} = \uurelof{\tau'}$.

We now turn our attention to proving that $\csrcrelof{\tau} = \csrcrelof{\tau'}$. We
will prove that $\csrcrelof{\tau} \subseteq \csrcrelof{\tau'}$. From symmetry it then
follows that $\csrcrelof{\tau'}\subseteq\csrcrelof{\tau}$, and hence $\csrcrelof{\tau} =
\csrcrelof{\tau'}$.
Let us assume that the relation $\anevent\csrcrelof{\tau}\anevent'$ exists in
$\csrcrelof{\tau}$for some events $\anevent = (\ptid,\upd{\xvar},j)$ and
$\anevent = (\ptid',\ld{\xvar},j')$. We will prove that the same
relation $\anevent\csrcrelof{\tau'}\anevent'$ exists in $\csrcrelof{\tau'}$.
From the definition of $\csrcrelof{\tau}$ we have that $\roweof{\anevent'}
\cmplt{\tau} \anevent \cmplt{\tau} \anevent'$ and there is no update
$\anevent'' = (\ptid'',\upd{\xvar},i'')$ to the same memory location
such that $\anevent\cmplt{\tau}\anevent''\cmplt{\tau}\anevent'$. Since
$\anevent'$ is preceded in $\tau$ by at least one update to \xvar{},
there must be a store event $\anevent_w$ such that
$\anevent_w\srcrelof{\tau}\anevent'$ in $\tau$. From the definition of
$\srcrelof{\tau}$ we have that $\anevent_w$ is the maximal event
$(\ptid'',\st{\xvar},j'')$ with respect to $\strelof{\tau}^*$ such that either
$\stupdof{\anevent_w}\cmplt{\tau}\anevent'$ or
$\anevent_w\porelof{\tau}^*\anevent'$.
If $\anevent_w\porelof{\tau}^*\anevent'$, then $\stupdof{\anevent_w} =
\roweof{\anevent'}$. But then the maximality of $\anevent_w$
contradicts $\roweof{\anevent'} \cmplt{\tau} \anevent \cmplt{\tau}
\anevent'$. Hence we have
$\stupdof{\anevent_w}\cmplt{\tau}\anevent'$. Maximality of
$\anevent_w$ now gives that $\stupdof{\anevent_w} = \anevent$.
Since $\srcrelof{\tau} = \srcrelof{\tau'}$ we have that in $\tau'$ also
$\anevent_w\srcrelof{\tau'}\anevent'$. From the definition of $\srcrelof{\tau'}$ and
$\neg(\anevent_w\porelof{\tau'}^*\anevent')$ we know that
$\stupdof{\anevent_w}$ is the store-order-maximal update to $\xvar$
that precedes $\anevent'$ in $\tau'$. Since the store order is the
same for $\tau$ and $\tau'$ we have
$\roweof{\anevent'}\cmplt{\tau'}\anevent$. But then
$\anevent = \stupdof{\anevent_w}$ satisfies the criteria for
$\anevent\csrcrelof{\tau'}\anevent'$.

Finally, we will show that $\ccfrelof{\tau} = \ccfrelof{\tau'}$. Similarly to the
proof for $\csrcrelof{\tau} = \csrcrelof{\tau'}$, it suffices here to show that
$\ccfrelof{\tau} \subseteq\ccfrelof{\tau'}$. Assume therefore that
$\anevent_r\ccfrelof{\tau}\anevent_u$ for some events $\anevent_r =
(\ptid,\ld{\xvar},j)$, $\anevent_u = (\ptid',\upd{\xvar},j')$. We will
show that $\anevent_r\ccfrelof{\tau'}\anevent_u$.
The definition of $\ccfrelof{\tau}$ gives that $\anevent_u$ is the first
(w.r.t. $\cmplt{\tau}$) event $\anevent$ of the form
$(\_,\upd{\xvar},\_)$ such that both $\anevent_r\cmplt{\tau}\anevent$
and $\roweof{\anevent_r}\cmplt{\tau}\anevent$.
Let $\anevent_w$ be the store event such that $\stupdof{\anevent_w} =
\anevent_u$.
We will split the proof in cases depending on whether or not there
exists a source event for $\anevent_r$ in the Shasha-Snir traces.

Assume therefore first (i) that there is no event $\anevent_{src}$
such that $\anevent_{src}\srcrelof{\tau}\anevent_r$.
Then there is no update to \xvar{} that precedes $\anevent_r$ in
$\cmplt{\tau}$. Furthermore $\roweof{\anevent_r} = \evtzero$. This tells us
that $\anevent_w$ has no predecessor in $\strelof{\tau}$.
Since $\strelof{\tau} = \strelof{\tau'}$, we also have that $\anevent_w$ has no
predecessor in $\strelof{\tau'}$. Furthermore, since $\anevent_r$ has no
source event in $\tau'$, it must be the case that
$\anevent_r\cmplt{\tau'}\anevent_u$. But then, $\anevent_u$ is the
first update event in $\tau'$ which is after both $\anevent_r$ and
$\roweof{\anevent_r}$. And so we have $\anevent_r\ccfrelof{\tau'}\anevent_u$.

Next assume (ii) that there is an event $\anevent_{src}$ with
$\anevent_{src}\srcrelof{\tau}\anevent_r$ and that $\tidof{\anevent_{src}} =
\tidof{\anevent_r}$.
Then it must be the case that $\stupdof{\anevent_{src}} =
\roweof{\anevent_r}$. Since $\srcrelof{\tau} = \srcrelof{\tau'}$, we have
that $\anevent_{src}\srcrelof{\tau'}\anevent_r$.
There can be no update event $\anevent$ to the same memory location
\xvar{} such that
$\roweof{\anevent_r}\cmplt{\tau}\anevent\cmplt{\tau}\anevent_r$. If
there were such an $\anevent$, then $\anevent_{src}$ wouldn't be the
source of $\anevent_r$. The same argument goes in $\tau'$.
This tells us that $\anevent_u$ is the immediate store order successor
of $\roweof{\anevent_r}$, i.e., $\roweof{\anevent_r}\uurelof{\tau}\anevent_u$ and
$\anevent_{src}\strelof{\tau}\anevent_w$. Since $\uurelof{\tau} = \uurelof{\tau'}$, we have
$\roweof{\anevent_r}\uurelof{\tau'}\anevent_u$. Hence $\anevent_u$ is the first
update event which succeeds both $\anevent_r$ and $\roweof{\anevent_r}$ in
$\cmplt{\tau'}$. Thus $\anevent_r\ccfrelof{\tau'}\anevent_u$.

Lastly, we assume (iii) that there is an event $\anevent_{src}$ such
that $\anevent_{src}\srcrelof{\tau}\anevent_r$ and that $\tidof{\anevent_{src}}
\neq \tidof{\anevent_r}$.
Then it is the case in $\tau$ that
$\roweof{\anevent_r}\cmplt{\tau}\stupdof{\anevent_{src}}\cmplt{\tau}\anevent_r$. And
there is no update event $\anevent$ to \xvar{} such that
$\stupdof{\anevent_{src}}\cmplt{\tau}\anevent\cmplt{\tau}\anevent_r$. The
same holds in $\tau'$. Since $\anevent_u$ is the first update to
\xvar{} after $\anevent_r$ in $\tau$, this means that we have
$\stupdof{\anevent_{src}}\uurelof{\tau}\anevent_u$. We have $\uurelof{\tau} =
\uurelof{\tau'}$, so $\stupdof{\anevent_{src}}\uurelof{\tau'}\anevent_u$.
Now it must be the case that
$\anevent_r\cmplt{\tau'}\anevent_u$. Otherwise, $\anevent_{src}$
wouldn't be the source of $\anevent_r$ in $\tau'$, and we know
$\anevent_{src}\srcrelof{\tau'}\anevent_r$. Hence $\anevent_u$ is an update
event that succeeds both $\anevent_r$ and $\roweof{\anevent_r}$ in
$\cmplt{\tau'}$. It remains to show that it is the first such
update. Suppose $\anevent \neq \anevent_u$ is an update event to
\xvar{} such that
$\anevent_r\cmplt{\tau}\anevent\cmplt{\tau}\anevent_u$. Then it would
be the case that
$\stupdof{\anevent_{src}}\cmplt{\tau'}\anevent\cmplt{\tau'}\anevent_u$. But
this would contradict
$\stupdof{\anevent_{src}}\uurelof{\tau'}\anevent_u$. Thus we have
$\anevent_r\cfrelof{\tau'}\anevent_u$.

This concludes the proof of $\chronof{\tau} = \chronof{\tau'}$.

\paragraph{Proof of Lemma~\ref{lemma:equiv:ss:chrono:left}}

Let two completed executions $\tau$ and $\tau'$ be given. Let\\
$\tsotraceof{\tau} = \langle
V_{SS},\porelof{\tau}\cup\strelof{\tau}\cup\srcrelof{\tau}\cup\cfrelof{\tau}\rangle$ and\\
$\tsotraceof{\tau'} = \langle V_{SS}',
\porelof{\tau'}\cup\strelof{\tau'}\cup\srcrelof{\tau'}\cup\cfrelof{\tau'}\rangle$ and\\
$\chronof{\tau} = \langle V_C,\porelof{\tau}\cup\surelof{\tau}\cup\uurelof{\tau}\cup\csrcrelof{\tau}\cup\ccfrelof{\tau}\cup\ufrelof{\tau}\rangle$ and \\
$\chronof{\tau'} = \langle V'_C,\porelof{\tau'}\cup\surelof{\tau'}\cup\uurelof{\tau'}\cup\csrcrelof{\tau'}\cup\ccfrelof{\tau'}\cup\ufrelof{\tau'}\rangle$.\\
Furthermore, assume that $\chronof{\tau} = \chronof{\tau'}$.

We will prove that $\tsotraceof{\tau} = \tsotraceof{\tau'}$. We know
that $V_{SS}$ (respectively $V'_{SS}$) is precisely the non-updates
of $V_C$ (respectively $V'_C$). Since $V_C = V'_C$ we have $V_{SS} =
V'_{SS}$.

For the relations $\porelof{\tau}$ and $\strelof{\tau}$, a reasoning analogue to that
in the $\Rightarrow$ direction gives that $\porelof{\tau} = \porelof{\tau'}$ and
$\strelof{\tau} = \strelof{\tau'}$.

We will show that $\srcrelof{\tau}\subseteq\srcrelof{\tau'}$. Symmetry then gives
$\srcrelof{\tau'}\subseteq\srcrelof{\tau}$, and hence $\srcrelof{\tau} = \srcrelof{\tau'}$.
Assume therefore that $\anevent_w\srcrelof{\tau}\anevent_r$ holds for
some events $\anevent_w = (\ptid,\st{\xvar},j)$ and $\anevent_r =
(\ptid',\ld{\xvar},j')$.
Then by the definition of $\srcrelof{\tau}$ we have that $\anevent_w$ is the
maximal event $\anevent = (\ptid'',\st{\xvar},j'')$ with respect to
$\strelof{\tau}^*$ such that either $\stupdof{\anevent}\cmplt{\tau}\anevent_r$
or $\anevent\porelof{\tau}^*\anevent_r$.
We will separate the proof by cases: either $\tidof{\anevent_w}
= \tidof{\anevent_r}$ or $\tidof{\anevent_w} \neq \tidof{\anevent_r}$.

Assume first (i) that $\tidof{\anevent_w} = \tidof{\anevent_r}$. Then
it holds that $\anevent_w\porelof{\tau}^*\anevent_r$, since the events must be
program ordered, and the other direction implies $\anevent_r
\cmplt{\tau} \stupdof{\anevent_w}$. Program order is the same in
$\tau'$ as in $\tau$, so we also have
$\anevent_w\porelof{\tau'}^*\anevent_r$. It remains to show that $\anevent_w$
is maximal in $\tau'$. First we conclude that there can be no store
event $\anevent$ such that $\anevent_w\strelof{\tau'}\anevent$ and
$\anevent\porelof{\tau'}^*\anevent_r$. This is because both the program order
and the store order are the same in $\tau'$ as in $\tau$, and hence
such an event $\anevent$ would contradict the assumed maximality of
$\anevent_w$ w.r.t. $\tau$. As a corollary we have
$\roweof{\anevent_r} = \stupdof{\anevent_w}$.
Next we need to conclude that there is no event $\anevent$ such that
$\anevent_w\strelof{\tau'}\anevent$ and
$\stupdof{\anevent}\cmplt{\tau'}\anevent_r$.
We know that there is no such event in $\tau$: i.e., there is no event
$\anevent$ such that $\anevent_w\strelof{\tau}\anevent$ and
$\stupdof{\anevent}\cmplt{\tau}\anevent_r$. Hence by the definition of
$\csrcrelof{\tau}$ there is no event $\anevent_{src}^C$ which is source
related with $\anevent_r$ in the chronological trace:
$\anevent_{src}^C\csrcrelof{\tau}\anevent_r$. Since $\csrcrelof{\tau} = \csrcrelof{\tau'}$,
the same holds in $\tau'$. Now if there were an event such as
$\anevent$ in $\tau'$, then $\anevent_r$ would have a source according
to $\csrcrelof{\tau'}$. This is a contradiction, and so there can be no such
$\anevent$ in $\tau'$.
Hence, $\anevent_w$ is the maximal store event w.r.t. $\strelof{\tau'}^*$
which is either updated $\cmplt{\tau'}$-before $\anevent_r$ or program
order-before $\anevent_r$.
That concludes the proof for the case that $\tidof{\anevent_w} =
\tidof{\anevent_r}$.

Next assume (ii) that $\tidof{\anevent_w} \neq
\tidof{\anevent_r}$. Clearly $\anevent_w$ is not program ordered with
$\anevent_r$. Hence the definition of $\srcrelof{\tau}$ gives that
$\stupdof{\anevent_w}\cmplt{\tau}\anevent_r$.
The maximality of $\anevent_w$ gives that $\roweof{\anevent_r}
\cmplt{\tau} \stupdof{\anevent_w}$, and that there is no update event
$\anevent = (\ptid'',\upd{\xvar},j'')$ such that
$\stupdof{\anevent_w}\cmplt{\tau}\anevent\cmplt{\tau}\anevent_r$. Then
we have
$\stupdof{\anevent_w}\csrcrelof{\tau}\anevent_r$ by the definition of $\csrcrelof{\tau}$. By $\csrcrelof{\tau} = \csrcrelof{\tau'}$ we
also have $\stupdof{\anevent_w}\csrcrelof{\tau'}\anevent_r$.
By the definition of $\csrcrelof{\tau'}$ we now have that $\anevent_w$ is the
greatest (w.r.t. $\cmplt{\tau'}$) store event with
$\stupdof{\anevent_w}\cmplt{\tau'}\anevent_r$. We also have that
$\roweof{\anevent_r}\cmplt{\tau'}\stupdof{\anevent_w}$. Since there
can be no event $\anevent = (\_,\st{\xvar},\_)$ such that
$\anevent\porelof{\tau'}^*\anevent_r$ and
$\roweof{\anevent_r}\cmplt{\tau'}\stupdof{\anevent}$, we have that
$\anevent_w$ is the maximal event $\anevent =
(\_,\st{\xvar},\_)$ with respect to $\strelof{\tau}^*$ such that either
$\stupdof{\anevent}\cmplt{\tau'}\anevent_r$ or
$\anevent\porelof{\tau'}^*\anevent_r$.
Hence $\anevent_w\srcrelof{\tau'}\anevent_r$.
This concludes the proof for $\srcrelof{\tau} = \srcrelof{\tau'}$.

Since $\cfrelof{\tau}$ (respectively $\cfrelof{\tau'}$) is entirely determined by
$\srcrelof{\tau}$ and $\strelof{\tau}$ (respectively $\srcrelof{\tau'}$ and $\strelof{\tau'}$), and we
know that $\srcrelof{\tau} = \srcrelof{\tau'}$ and $\strelof{\tau} = \strelof{\tau'}$, we immediately
get that $\cfrelof{\tau} = \cfrelof{\tau'}$. This concludes the proof.

\paragraph{Proof of Theorem~\ref{thm:equivalence}}

The theorem follows directly from
Lemmas~\ref{lemma:equiv:ss:chrono:right}
and \ref{lemma:equiv:ss:chrono:left}.

\section{DPOR for TSO} \label{app:algorithm}

In this appendix, we establish the correctness of
Theorem~\ref{thm:correctness}, which states
that the DPOR algorithms {\em Source-DPOR} and
{\em Optimal-DPOR} of~\cite{abdulla2014optimal}, when based on the
happens-before relation induced by chronological traces, explore at least
one execution per equivalence class induced by Shasha-Snir traces.
Theorem~\ref{thm:correctness} also states that
{\em Optimal-DPOR} explores
exactly one execution per equivalence class.
We also provide more detail on how a DPOR algorithm, such as
{\em Source-DPOR} of~\cite{abdulla2014optimal}, can be used for SMC
on programs running under TSO by computing chronological traces on the fly.

Correctness of
{\em Source-DPOR} also implies that several other DPOR algorithms, e.g.,
in~\cite{FG:dpor,SKH:acsd12} would be correct if based on chronological traces.
This is because these algorithms are subsumed by {\em Source-DPOR} in the
sense that the set of executions that are explored by these algorithms in
some particular analysis includes the set of executions that could be
explored by {\em Source-DPOR} in some analysis.

\begin{theorem}
{\bf (Correctness of DPOR algorithms)}
The algorithms {\em Source-DPOR} and {\em Optimal-DPOR}
of~\cite{abdulla2014optimal}, based on the
happens-before relation induced by chronological traces, explore at least
one execution per equivalence class induced by Shasha-Snir traces.
Moreover, {\em Optimal-DPOR} explores exactly one execution per equivalence
class.
\end{theorem}

\paragraph{Proof}
The proof of Theorem~\ref{thm:correctness} mainly uses the correctness
of {\em Source-DPOR}, which is proven in~\cite{abdulla2014optimal}.
More precisely, in~\cite{abdulla2014optimal} it is proven that
{\em Source-DPOR} is correct whenever it is based on an assignment of
happens-before relations to executions, which is {\em valid}.
An assignment of happens-before relations $\happensbefore{\exseq}$
to executions $\exseq$ is valid if
it satisfies the following natural properties
(from ~\cite{abdulla2014optimal}).
\begin{enumerate}
\item \label{cond:1}
$\happensbefore{\exseq}$ is a partial order on the events in
$\exseq$, which is included in $\cmplt{\exseq}$,
\item \label{cond:2}
the events of each thread are totally ordered by
$\happensbefore{\exseq}$,
\item \label{cond:3}
if $\exseq'$ is a prefix of $\exseq$,
  then $\happensbefore{\exseq}$ and
  $\happensbefore{\exseq'}$ are the same on $\exseq'$.
\item \label{cond:4}
the assignment of happens-before relations to executions
partitions the set of executions into equivalence classes;
i.e., if $\exseq'$ is a linearization of the happens-before
relation on $\exseq$, then $\exseq'$ is assigned the same
happens-before relation as $\exseq$;
we use $\mtequiv$ to denote the corresponding equivalence relation,
\item \label{cond:5}
whenever $\exseq$ and $\exseq'$ are equivalent then they
end up in the same global program state,
\item \label{cond:6}
  for any sequences $\exseq$, $\exseq'$ and $\exseq''$, such that
$\exseq\cdot{}\exseq''$ is an execution, we have $\exseq
  \mtequiv \exseq'$ if and only if
$\exseq\cdot{}\exseq'' \mtequiv \exseq'\cdot{}\exseq''$, and
\item \label{cond:7}
if $\exseq\cdot{}(p,i,j)$ is an execution, whose last event is performed by
thread $p$,
and $q$, $r$ are different threads, such that $(p,i,j)$ would
``happen before'' a subsequent event by $r$ but not a subsequent event by $q$,
then $(p,i,j)$ would also ``happen before'' $(r,i'',j'')$ in the execution
$\exseq\cdot{}(p,i,j)\cdot{}(q,i',j')\cdot{}(r,i'',j'')$.
\end{enumerate}
A consequence of these definitions is that
that if $\event$ and~$\event'$ are two consecutive events
in $\exseq$ with $\event \not\happensbefore{\exseq} \event'$, then
$\event$ and~$\event'$  can be swapped without affecting
the (global) state after the two events.

The theorem can now be proven by establishing that the happens-before
assignment induced by chronological traces is valid. Conditions
\ref{cond:1}, \ref{cond:2}, \ref{cond:3}, and \ref{cond:6}
follow straight-forwardly from definitions
Condition \ref{cond:4} follows by observing that changing the order between
non-related events does not affect the definition of the chronological
trace.
Condition \ref{cond:5} follows by observing that the chronological trace
captures all dependences that are needed for determining which
values are read and written by loads and stores.
Finally, Condition \ref{cond:7} follows by noting that an arrow
between $(p,i,j)$ and $(r,i'',j'')$ in a chronological trace
cannot be removed by inserting an event that is independent with $p$.
This concludes the proof of Theorem~\ref{thm:correctness}.
\qed

\medskip
We next provide more details on the computation of the
happens-before relation induced by chronological traces.

The happens-before relation $\ctrel$ is computed using
vector clocks, while taking the particular structure of
chronological traces into account.
The main difference from computing happens-before relations for sequentially
consistent executions (see, e.g.,~\cite{SKH:acsd12}) is that
load events which get their value by store forwarding are not
immediately synchronized with the vector clock of the memory
location. Instead the load is associated with the store buffer entry
from which it got its value. The load is then synchronized with the
memory location at the time when the store buffer entry is updated to
memory.

Formally, we introduce auxiliary configurations, and define the
semantics of instructions over them. When exploring an execution, all
instructions will be applied simultaneously to the TSO system configuration
(as described in Section~\ref{sec:semantics}) and the auxiliary
configuration.
Below we need vector clocks.
A {\em vector clock} is a function $C :
(\tids\cup\auxtids)\mapsto\nat$.
The intuition is that $C$ captures a set of observed events.
For every thread \ptid, the first $C(\ptid)$ events by
\ptid have been observed.
We let $\vecclocks = ((\tids\cup\auxtids)\mapsto\nat)$ denote the set
of vector clocks.

An {\em auxiliary configuration} is a triple
$(\mathcal{C},\mathcal{B},\mathcal{M})$, where
\begin{description}
\item[$\mathcal{C}$] $: (\tids\cup\auxtids\cup\events\cup\{\perp\})\mapsto\vecclocks$\\
 maps each (real or auxiliary) thread identifier $\ptid$ to
a vector clock representing
which parts of the execution have been seen by $\ptid$.
Also, $\mathcal{C}$ maps each event $\anevent$ to the value of
$\mathcal{C}(\tidof{\anevent})$ at the time immediately after
executing $\anevent$. We fix that $\mathcal{C}(\perp) = (\lambda x
. 0)$ is a zeroed clock.
\item[$\mathcal{B}$] $: \tids\mapsto(\memlocs\times\events\times(\events\cup\{\perp\}))^*$\\
maps each real (not auxiliary) thread ID $\ptid$ to
a word of letters $(\xvar,\anevent_s,\anevent_l)$, each of which
keeps auxiliary state for the corresponding letter in the store buffer
$\bufstate(\ptid)$. Here $\xvar$ is the accessed memory location,
$\anevent_s$ is the store event that produced that letter,
and $\anevent_l$ is the latest buffer forwarded load event for which the letter
has been the source (if there is no such event then $\anevent_l = \perp$).
\item[$\mathcal{M}$] $: \memlocs\mapsto((\events\cup\{\perp\})\times 2^{\events})$\\
maps each memory location $\xvar$ to
a pair $(\anevent_u,E_l)$, where $\anevent_u$ is the latest update event that
accessed $\xvar$ (or $\perp$ if $\xvar$ has never been updated), and where
$E_l$ is a set which for each thread $\ptid$ that has read \xvar{} contains
the latest event of $\ptid$ that read the value of $\xvar$.
\end{description}
Initially all clocks in $\mathcal{C}$ are zeroed, all buffers in
$\mathcal{B}$ are empty, and for all memory locations $\xvar$ we have
$\mathcal{M}(\xvar) = (\perp,\emptyset)$.

The idea here is that as we execute memory accesses, we update the
vector clock of the executing thread to reflect which new events have
been observed.

For example, when we execute an update $\anevent_{\xvar}$
which corresponds to a buffer entry $(\xvar,\anevent_s,\anevent_l)$, we
look to the memory $\mathcal{M}(\xvar) = (\anevent_u,E_l)$.
We know that the update event is ordered after the previous update
$\anevent_u$, as well as the previous loads in $E_l$ and the store
event $\anevent_s$ which enabled the update $\anevent_x$.
We update the vector clock
$\mathcal{C}(\tidof{\anevent_{\xvar}})$ of the auxiliary
thread to include all these newly observed events.

The procedure for a load from memory is similar, except that we do not
observe previous loads. More interesting are loads that are satisfied
by buffer forwarding.
When we execute a buffer forwarded load $\anevent_l$ to \xvar{}, we do
not observe {\em any} new event, since the load was not able to reach
and synchronize with the memory.
Instead we save the load event with the buffer entry from which it
read its value.
When that entry is updated to memory, by the update event
$\roweof{\anevent_l}$, we move $\anevent_l$ to the set of loads that
have been observed by $\mathcal{M}(\xvar{})$.
By this scheme the load event $\anevent_l$ becomes available for
observation by precisely the update events which succeed
$\roweof{\anevent_l}$.
In the remainder of this section we will make this intuition formal.

We will also need some notation for dealing with vector clocks.
For two vector clocks $v,v'$ we use the notation $v+v'$ to
denote the vector clock $v''$ such that $v''(\ptid) =
max(v(\ptid),v'(\ptid))$ for all $\ptid$. For two vector
clocks $v,v'$ we say that $v \leq v'$ when $v(\ptid) \leq v'(\ptid)$
for all $\ptid$. We say that $v < v'$ if at
least one of the inequalities is strict.
For an event $\anevent$ and a set $E$ of events we define
$E\oplus\anevent = \{\anevent'\in E | \tidof{\anevent'} \neq
\tidof{\anevent}\}\cup\{\anevent\}$,
i.e. $E\oplus\anevent$ is $E$ where $\anevent$ replaces the previous
event $\anevent'\in E$ s.t. $\tidof{\anevent'} = \tidof{\anevent}$.
We use the shorthand
$f[x_0,x_1,\cdots,x_n\hookleftarrow v]$ to denote
$f[x_0\hookleftarrow v][x_1\hookleftarrow v]\cdots[x_n\hookleftarrow v]$, i.e.,
an assignment of the same value to multiple function arguments.

For two arbitrary auxiliary configurations
$c=(\mathcal{C},\mathcal{B},\mathcal{M})$ and
$c'=(\mathcal{C}',\mathcal{B}',\mathcal{M}')$ we now define the
transition relation $c\xrightarrow{\ptid}c'$ depending on the next
instruction $op$ of $\ptid$.
We let $j = \mathcal{C}(\ptid)(\ptid)+1$ be the index of the next
event for $\ptid$ and $C_p = \mathcal{C}(\ptid)[\ptid\hookleftarrow j]$
be the corresponding clock and $\anevent = (\ptid,op,j)$ be the event
itself.

At the same time as we compute the next auxiliary configuration, we
also compute the set $R(\anevent)$ of races $(\anevent_r,\anevent_r')$
such that $\anevent_r' = \anevent$ and $\anevent_r$ is some earlier
event.
Recall that the races are the pairs of events from different threads
related by $\uurel$, $\csrcrel$, or $\ccfrel$.

\paragraph{Store:}
If $op = \st{\xvar}$, then $c\xrightarrow{\ptid}c'$ iff $\mathcal{C}'
= \mathcal{C}[\ptid,\anevent\hookleftarrow C_p]$, and $\mathcal{M}' =
\mathcal{M}$, and $\mathcal{B'} =
\mathcal{B}[\ptid\hookleftarrow\mathcal{B}(\ptid)\cdot(\xvar,\anevent,\perp)]$. There
are no races: $R(\anevent) = \emptyset$.

\paragraph{Load from memory:}
If $op = \ld{\xvar}$ and there is no letter on the form
$(\xvar,\_,\_)$ in $\mathcal{B}(\ptid)$, then $c\xrightarrow{\ptid}c'$
iff $\mathcal{C}' = \mathcal{C}[\ptid,\anevent\hookleftarrow C_p']$,
and $\mathcal{B}' =
\mathcal{B}$ and $\mathcal{M}' =
\mathcal{M}[\xvar\hookleftarrow(\anevent_u,E\oplus\anevent)]$,
where $\mathcal{M}(\xvar) = (\anevent_u,E)$.
Here $C_p' = C_p+\mathcal{C}(\anevent_u)$ if $\anevent_u\neq\perp$ and
$\tidof{\anevent_u} \neq \updof{p}$, and $C_p' = C_p$ otherwise.
If $\mathcal{C}(\anevent_u) \not\leq C_p$ and $\tidof{\anevent_u} \neq
\updof{\ptid}$, then we have $R(\anevent) = \{(\anevent_u,\anevent)\}$. Otherwise $R(\anevent) =
\emptyset$.

Intuitively, $\anevent$ is ordered after the last update $\anevent_u$
to \xvar{}, provided that $\anevent_u$ originated in a different
thread.

\paragraph{Load from buffer:}
If $op = \ld{\xvar}$ and $\mathcal{B}(\ptid) =
\mathcal{B}_0\cdot(\xvar,\anevent_u,\anevent_l)\cdot\mathcal{B}_1$ for
some $\mathcal{B}_0, \mathcal{B}_1, \anevent_u, \anevent_l$ with no
letters on the form $(\xvar,\_,\_)$ in $\mathcal{B}_1$, then
$c\xrightarrow{\ptid}c'$ iff $\mathcal{C}' =
\mathcal{C}[\ptid,\anevent\hookleftarrow C_p]$ and $\mathcal{B}' =
\mathcal{B}_0\cdot(\xvar,\anevent_u,\anevent)\cdot\mathcal{B}_1$ and
$\mathcal{M}' = \mathcal{M}$.

Notice that $\anevent$ replaces $\anevent_l$ in the store buffer
entry. There are no races: $R(\anevent) = \emptyset$.

\paragraph{Fence:}
If $op = \fence$ then $c\xrightarrow{\ptid}c'$ iff
$\mathcal{B}' = \mathcal{B}$ and $\mathcal{M}' = \mathcal{M}$ and
$\mathcal{C}' = \mathcal{C}[\ptid,\anevent\hookleftarrow
  C_p+\mathcal{C}(\updof{\ptid})]$.
There are no races: $R(\anevent) = \emptyset$.

Notice that the semantics of the fence, as defined in
Section~\ref{sec:semantics}, guarantee that $\mathcal{B}(\ptid) =
\varepsilon$. Hence the vector clock of the auxiliary thread
$\updof{\ptid}$ includes the clocks of all updates of $\updof{\ptid}$
corresponding to earlier stores.
So $\anevent$ will be ordered after all updates of $\updof{\ptid}$, as
prescribed by $\ufrel$ for chronological traces.

\paragraph{Update:}
If $op = \upd{\xvar}$ then $c\xrightarrow{\ptid}c'$ iff
$\mathcal{B} = (\xvar,\anevent_s,\anevent_r)\cdot\mathcal{B}'$ and
$\mathcal{C}' = \mathcal{C}[\ptid,\anevent\hookleftarrow C_p +
  \mathcal{C}(\anevent_s) + \mathcal{C}(\anevent_u) +
  \sum_{\anevent_l \in E \textrm{ s.t. } \updof{\tidof{\anevent_l}}
    \neq \ptid}\mathcal{C}(\anevent_l)]$
where $\mathcal{M}(\xvar) = (\anevent_u,E)$ and $\mathcal{M}' =
\mathcal{M}[\xvar\hookleftarrow(\anevent,E')]$. Here $E' = E$ if
$\anevent_r = \perp$, and $E' = E\oplus \anevent_r$ otherwise.
There is a race with every previous access to $\xvar$ from a different
thread:\\

\noindent
\begin{math}
R(\anevent) = \left\{\begin{array}{@{}c|c@{}}(\anevent',\anevent) &
\begin{array}{l}
\anevent'\in E\cup\{\anevent_u\} \wedge \anevent'\neq\perp \wedge \\
\tidof{\anevent'} \neq \ptid \wedge \updof{\tidof{\anevent'}} \neq \ptid \wedge \\
\mathcal{C}(\anevent')\not\leq C_p + \mathcal{C}(\anevent_s) \\
\end{array}
\end{array}
\right\}
\end{math}

\section{Adaptation for PSO}\label{sec:pso}

In this appendix, we show how our
techniques can be adapted to the PSO memory model with minor
changes. Before we see how to apply our methods to it,
we give an informal description of the PSO memory model.

\subsection{PSO Semantics}
\label{sec:pso:semantics}

PSO is a strictly more relaxed model than TSO. As described
previously, TSO allows reordering of stores with subsequent loads. PSO
allows the same reordering, but also allows the reordering of stores
with subsequent stores to different memory locations.

This behavior can be explained by an operational semantics similar to
the one described in Section~\ref{sec:semantics} for TSO, but where
each thread has a separate store buffer for each memory location.
Each store buffer is FIFO-ordered, so stores to the same memory
location by the same thread cannot be reordered. But there is no order
maintained between stores in different buffers, so stores by the same
thread to different locations may update in reversed order.

In Figure~\ref{fig:mp:prog} we give an example of a program where PSO
allows more behaviors than TSO. The execution in
Figure~\ref{fig:mp:cmp} shows how the stores by \ptid to \xvar and
\yvar update to memory in reversed order. This allows the thread
\qtid to read first $\yvar = 1$ then $\xvar = 0$, which would be
impossible both under SC and TSO.

In the operational semantics for PSO we introduce one auxiliary thread
$\updof{\ptid,\xvar}$ for each pair of a thread \ptid and a memory
location \xvar. Each such auxiliary thread is responsible for the
updates to \xvar by \ptid, similarly to how $\updof{\ptid}$ under
TSO is responsible for all updates of \ptid.

\begin{figure}
  \centering
  \begin{subfigure}[b]{.45\linewidth}
    \centering
    \begin{tabular}{@{}ll@{}}
      \multicolumn{1}{c}{\ptid} & \multicolumn{1}{c}{\qtid}\\
      \store{\xvar}{1} & \load{\rreg}{\yvar} \\
      \store{\yvar}{1} & \load{\sreg}{\xvar} \\
    \end{tabular}
    \caption{The mp idiom. Possible under PSO: $\rreg = 1$, $\sreg = 0$.}\label{fig:mp:prog}
  \end{subfigure}
  \rule{10pt}{0pt}
  \begin{subfigure}[b]{.45\linewidth}
    \centering
    \begin{tabular}{@{}l@{}}
      $(\ptid, \st{\xvar}, 1)$\\
      $(\ptid, \st{\yvar}, 2)$\\
      $(\updof{\ptid,\yvar},\upd{\yvar},1)$\\
      \pindent$(\qtid,\ld{\yvar},1)$\\
      \pindent$(\qtid,\ld{\xvar},2)$\\
      $(\updof{\ptid,\xvar},\upd{\xvar},1)$\\
    \end{tabular}
    \caption{An execution $\tau$ where finally $\rreg = 1$, $\sreg = 0$.}\label{fig:mp:cmp}
  \end{subfigure}

  \begin{subfigure}[b]{.9\linewidth}
    \centering
    \begin{tikzpicture}
      \node (wx) at (0,0) [] {%
        \small{\st{\xvar}}
      };
      \node (wy) at ($(wx.west) + (0,-1)$) [anchor=west] {%
        \small{\ld{\yvar}}
      };

      \node (ux) at ($(wy) + (2,0)$) [] {%
        \small{\upd{\xvar}}
      };
      \node (uy) at ($(wx) + (4,0)$) [] {%
        \small{\upd{\yvar}}
      };

      \node (ry) at (6,0) [] {%
        \small{\ld{\yvar}}
      };
      \node (rx) at ($(ry.west) + (0,-1)$) [anchor=west] {%
        \small{\ld{\xvar}}
      };

      \draw[->,line width=1pt] (wx) -- node [left] {\small{\textsf{po}}} (wx |- wy.north);
      \draw[->,line width=1pt] (ry) -- node [right] {\small{\textsf{po}}} (ry |- rx.north);
      \draw[->,line width=1pt] (wx) -- node [above] {\small{\textsf{su}}} (ux);
      \draw[->,line width=1pt] (wy) -- node [above] {\small{\textsf{su}}} (uy);
      \draw[->,line width=1pt] (uy) -- node [above] {\small{\textsf{src}}} (ry);
      \draw[->,line width=1pt] (rx) -- node [below] {\small{\textsf{cf}}} (ux);

      \node (p) at ($(wx) + (0,0.75)$) [] {\ptid};
      \node (q) at ($(ry) + (0,0.75)$) [] {\qtid};

      \node (px) at (p -| ux) [] {\updof{\ptid,\xvar}};
      \node (py) at (p -| uy) [] {\updof{\ptid,\yvar}};
      \tikzwrapfigbg
    \end{tikzpicture}
    \caption{The chronological trace of $\tau$ under PSO.}
    \label{fig:mp:chronotrace}
  \end{subfigure}

  \caption{A behavior allowed under PSO but not under TSO.}
  \label{fig:mp}
\end{figure}

\subsection{Chronological Traces for PSO}

The adaptation of chronological traces to PSO is straightforward. The
following simple adjustment suffices:
Since stores from the same thread \ptid{} to different memory
locations \xvar{} and \yvar{} are updated by different auxiliary
threads \updof{\ptid,\xvar} and \updof{\ptid,\yvar}, there is no
program order edge between the update events for different memory
locations under PSO.

A chronological trace for PSO is illustrated in
Figure~\ref{fig:mp:chronotrace}.
Notice that there is no program order edge from
$(\updof{\ptid,\xvar},\upd{\xvar},1)$ to
$(\updof{\ptid,\yvar},\upd{\yvar},1)$.
Had there been one, the trace would be cyclic.

\section{Chronological Traces Versus Xtop Objects}\label{sec:xtop}

In this appendix, we compare chronological traces and Xtop objects~\cite{AlKNT13}.

Our goal was to design a mechanism for representing executions under
relaxed memory models that would allow a stateless model checking
technique to not explore more that one execution per Shasha-Snir trace.
In fact, chronological traces allowed us to achieve this goal, since
they are acyclic (in contrast to vanilla Shasha-Snir) but still
correspond one-to-one to Shasha-Snir traces.
This is not the case for Xtop objects~\cite{AlKNT13},
since two Xtop objects may map to the same Shasha-Snir trace.
In fact, chronological traces and Xtop objects are different.
This is because an Xtop object includes information
about which event pairs are reordered (delayed), while a chronological
trace does not. As an illustrative example, consider the SB idiom. A
chronological trace is given in \figurename~\ref{fig:dekker:ctrace},
and a corresponding Xtop object is given in~\cite[\figurename~4(b)]{AlKNT13}.
Notice that the edges $f(b)-de->f(a)$ and $f(c)-se->f(d) $ in the
Xtop object specify that $a$ and $b$ are reordered, while $c$ and $d$ execute
in program order. This choice is not imposed by the chronological
trace. Indeed, there is a different Xtop object where $c$ and $d$ are
reordered instead of $a$ and $b$. Both of these Xtop objects correspond to
the same chronological trace and to the same Shasha-Snir trace
(\figurename~\ref{fig:dekker:cmp:trace}).
Hence a POR technique that explores precisely all Xtop objects would
unnecessarily explore more objects than our technique.

\end{document}